# Integrated Information Theory: A Consciousness-First Approach to What Exists

Giulio Tononi[1] and Melanie Boly[2]

1. Dept. of Psychiatry, University of Wisconsin-Madison, WI, USA
2. Dept. of Neurology, University of Wisconsin-Madison, WI, USA

Correspondence to gtononi@wisc.edu

We thank the many lab members and colleagues who have contributed to the development of IIT over the years. This work was made possible through support from the Tiny Blue Dot Foundation, the Fries Family Fellowship, and the Templeton World Charity Foundation. The opinions expressed in this publication are those of the authors and do not necessarily reflect the views of these funding institutions.

G.T. is Board Member and has a financial interest in Intrinsic Powers Inc. G.T. holds patent rights along with the Wisconsin Alumni Research Foundation: patent No. US 8,457,731 B2, and patent applications Wisconsin Alumni Research Foundation No. P220191, P220192. No AI was involved in any aspect of this overview.

To appear in L. Melloni & U. Olcese (Eds.), The Scientific Study of Consciousness: Experimental and Theoretical Approaches. Springer-Nature (forthcoming).

*Summary*. This overview of integrated information theory (IIT) emphasizes IIT's 'consciousness-first' approach to what exists. Consciousness demonstrates to each of us that something exists—experience—and reveals its essential properties—the axioms of phenomenal existence. IIT formulates these properties operationally, yielding the postulates of physical existence. To exist intrinsically or absolutely, an entity must have cause-effect power upon itself, in a specific, unitary, definite and structured manner. IIT's explanatory identity claims that an entity's cause-effect structure accounts for all properties of an experience—essential and accidental—with no additional ingredients. These include the feeling of spatial extendedness, temporal flow, of objects binding general concepts with particular configurations of features, and of qualia such as colors and sounds. IIT's intrinsic ontology has implications for understanding meaning, perception, and free will, for assessing consciousness in patients, infants, other species, and artifacts, and for reassessing our place in nature.

## Introduction: experience as intrinsic existence

When I am conscious, there is something rather than nothing—'there is something it is like’ to exist. When I lose consciousness, as far as I am concerned, everything vanishes. Consciousness can thus be considered as *intrinsic*, *absolute*, or genuine existence [i]—the only existence worth being—because without it there would be nothing at all. Imagine a universe without consciousness. In what sense would it genuinely exist, if it would not exist for anyone? As Erwin Schrödinger put it, a world without consciousness would have remained “a play before empty benches, not existing for anybody, thus quite properly speaking not existing” (Schrödinger, 1992).

It may seem odd to begin this overview of integrated information theory (IIT) with a statement about what it means to exist. The usual way of approaching consciousness scientifically is to presuppose all we have learned about the universe through physics, chemistry, biology, and neuroscience. We can then finally ask how some part of the ‘physical’ universe—namely the brain—would ‘give rise’ to experience, or how consciousness might ‘emerge’ from some special kind of neural process or computation. It is also generally assumed that, with the objective tools of science, we can only explain and predict objective properties. To wit, neuroscience has made great progress in understanding how the brain is organized and carries out various functions, including those we typically perform when we are conscious, like recognizing people and objects, talking, pursuing goals, and making decisions. And yet, if it weren’t for the fact that each of us is conscious, there would be no reason to suspect that, when our brain does what it does, doing is accompanied by being—by a subject who experiences sights, sounds, and thoughts.

IIT takes the opposite approach. It presupposes consciousness, which is subjective, and aims to account for its presence and properties—for what it *is*, rather than what it *does*—in objective terms. It goes from phenomenology to physics—from what exists subjectively to an objective account of it—rather than trying to conjure phenomenology out of physics.

What follows is an overview of IIT and its implications, beginning with the foundations of its consciousness-first approach. To help navigate the text, here is an outline of its sections.

*Axioms of phenomenal existence*. Experience is not just proof that something exists, rather than nothing. It also demonstrates the essential properties of existence—those that are true of every conceivable experience. These are codified as IIT’s axioms of phenomenal existence: intrinsicality (experience exists for itself), information (it is specific), integration (it is irreducible), exclusion (it is definite), and composition (it is structured).

*Postulates of physical existence*. The next step is to formulate the axioms of phenomenal existence as postulates of physical existence. In IIT, ‘physical’ is understood in strictly operational terms as cause-effect power—the ability to take and make a difference from the perspective of a conscious subject. IIT’s postulates express the essential properties of experience as causal properties of a substrate, where a substate is understood as a set of units that can be observed and manipulated. This step introduces critical quantities such as intrinsic information and integrated information, which measures the irreducibility of a substrate’s cause-effect power. The operational formulation of phenomenal properties as causal properties is critical because it allows for the objective assessment of subjective existence.

*Explanatory identity*. The first four postulates (intrinsicality, information, integration, and exclusion) permit the identification of complexes—sets of units in their current state that are maximally irreducible. The units of a complex, called intrinsic units, must also satisfy the postulates. By ‘unfolding’ the cause-effect power of the complex’s subsets according to IIT’s fifth postulate (composition), one obtains the complex’s cause-effect structure, composed of causal distinctions and relations. IIT’s explanatory identity claims that the complex’s cause-effect structure accounts for all properties of an experience, essential and accidental, with no additional ingredients.

*Empirical validation*. The next section examines the empirical validity of IIT—what it explains and predicts—as established in ourselves (being intrinsic, its attribution to others is necessarily a matter of inference). IIT’s first four postulates have been employed to account for basic facts about the presence of consciousness, such as its dependence on certain parts of the brain and not others, its vanishing during dreamless sleep, and its return during dreaming sleep. They have led to several testable predictions and to practical methods for assessing consciousness in unresponsive subjects. IIT’s composition postulate can be used to account for the quality of consciousness. It leads to various predictions about the anatomical organization of brain substrates whose unfolded cause-effect structure can account for qualities such as spatial extendedness and temporal flow. Ongoing work aims at accounting for other contents of experience such as objects, which bind general concepts with particular configurations of features, and for qualia (in the narrow sense), such as colors and sounds.

*Intrinsic meaning, perception, and matching*. If we trust IIT’s construct validity and its empirical validation (so far), several implications follow. A direct consequence of IIT is that the meaning of an experience is the same as its feeling—’the meaning is the feeling.’ Every content of experience corresponds to a sub-structure within a cause-effect structure and is thus fully intrinsic—regardless of whether it is triggered by perception, imagined, or dreamt. Accordingly, IIT conceptualizes perception as the triggering of intrinsic meanings by a stimulus—as interpretation rather than as information processing or representation. The relationship between contents of experience and the environment is also conceptualized as one of matching, rather than reference or inference. IIT’s notion of integrated information as meaning is orthogonal to Shannon’s notion of information as message. A corollary of IIT’s notion is that the communication of integrated information as meaning requires the triggering of similar cause-effect structures, rather than the mere transfer of messages.

*Richness of experience*. Within cognitive science and psychology, it is regularly claimed that we experience much less than we think. According to IIT, the opposite is true: to feel the way it does, every experience, even that of pure darkness and silence, must be an immensely rich phenomenal structure. In line with IIT’s explanatory identity, the corresponding cause-effect structure must be equally rich. This is expected given the combinatorics of composition, as the number of distinctions and relations specified by a large complex is hyper-astronomical.

*Intrinsic powers ontology*. IIT characterizes the properties of existence based on those of experience—the only intrinsic or absolute existence. By operationalizing existence as cause-effect power, IIT configures what might be called an ‘intrinsic powers ontology.’ The universe, which is conceived operationally as ‘cause-effect power all the way down,’ condenses into a number of non-overlapping ‘intrinsic entities’—complexes with their associated cause-effect structure—which satisfy IIT's postulates. Nevertheless, the postulates also offer the tools for characterizing 'extrinsic entities,' which are highly integrated without being absolute maxima of integrated information. Many everyday 'things,' such as apples or tables, and large 'things,' such as planets or stars, likely qualify as extrinsic entities. However, between intrinsic and extrinsic existence passes the 'great divide of being'—existing for oneself, thus existing absolutely, vs. merely existing for something else, relative to it. More generally, the postulates allow for considering multiple or 'emerging' levels of existence, as long as it is clear that only intrinsic entities—i.e. conscious ones—genuinely exist. IIT's intrinsic powers ontology also implies that intrinsic entities exist as structures 'here and now,' rather than as dynamical processes ('being is not happening'). Because experience is a structure, it is also not a function or a computation ('being is not doing'). Finally, contents of experience, whether concrete like a face or an apple or abstract like beauty or justice, exist intrinsically—they genuinely exist (as sub-structures), whether or not they have a clear extrinsic referent. And they exist within each individual mind, rather than in a Platonic heaven.

*Free will and responsibility*. Among the contents of experience that genuinely exist are questions and answers, problems and solutions, as well as alternatives, values, goals, reasons, and decisions. On this basis, IIT's intrinsic ontology leads to the conclusion that we have genuine free will. And because 'only what exists can cause,' we, and not our neurons, are the cause of our actions (as can be assessed through IIT's analysis of actual causation). Moreover, the more we exercise free will in self-changing actions, the more we acquire genuine responsibility.

*Consciousness in patients, infants, other species, and artifacts*. Because it defines the requirements for subjective existence in objective terms, IIT should ideally provide a principled answer to questions about the presence and quality of consciousness beyond adult humans. In every case, the answer depends on the properties of the substrate—primarily its 'anatomy' and 'physiology'—rather than on intelligence and cognitive abilities. This applies to patients with severe disorders of consciousness, to developing humans, to species different from us, and to artifacts endowed with artificial intelligence (AI). In most cases, our knowledge of various substrates and their ability to support complexes with rich cause-effect structures is still inadequate to provide a reliable answer. In the case of present-day computers, however, IIT's postulates can already be used to prove that, even if they were functionally equivalent to us, they would not be phenomenally equivalent. In other words, they may soon do everything we do without experiencing anything ('doing without being').

*Consciousness and our place in nature*. At least since Galileo, science has adopted the extrinsic perspective on what exists. If we take the 'physical world' as ontologically primary, consciousness seems to belong to a different domain of existence. This has led to absurd consequences, from the denial and denunciation of consciousness as a meaningless word, to its dismissal as folk psychology, its demotion to illusion, its diminishment to a few tokens of reportable content, or its deportation outside the realm of science. Even more consequentially, science's adoption of the extrinsic perspective on existence has led to several ontological displacements—that our place in nature is peripheral and ephemeral, our inner life nothing but a bag of neural computations, our origin an evolutionary game of chance and circumstance, and our worth soon to be superseded by AI. If one takes the intrinsic perspective, however, it becomes possible to formulate a unified scientific ontology—one that starts from consciousness but can potentially account, with the same objective tools, for both extrinsic and intrinsic existence. Moreover, as will be briefly argued at the end of this overview, the displacements concerning our place in nature and the significance of our inner life are reversed.

Despite its length, this overview can only provide a broad picture of IIT and its implications, rather than an in-depth, technical exposition. Moreover, citations had to be kept to a minimum. The first section, which provides the necessary foundations and nomenclature of IIT, is especially dense and made harder by the lack of examples and explanations. Those who wish to explore IIT in greater detail should consult the original publications and the IIT Wiki (2024). On the other hand, this overview intentionally does not shy away from ontology, nor should it, because consciousness is existence. All kinds of misunderstandings, not to mention scientific blind spots, occur if one does not acknowledge this fact. Moreover, this overview argues that, as a scientific, testable theory of consciousness as intrinsic existence, IIT is also a testable *scientific ontology*, one that carries several metaphysical implications.

## The foundations of IIT

As we have seen, IIT starts from experience itself—phenomenal or intrinsic existence. It proceeds to identify the properties that are irrefutably true of every conceivable experience—the axioms of phenomenal existence. It then formulates the axioms in operational terms, yielding IIT's postulates of physical existence. Finally, it proposes an explanatory identity between experiences and cause-effect structures supported by a substrate's intrinsic units.

### IIT's axioms: the essential properties of experience

IIT's $0^{th}$ axiom, *existence*, can be expressed as follows: experiencing is *what it is like to be*. In other words, if there is experience, something exists, immediately and irrefutably. I can conceive of not experiencing anything, but then there would not be anything it is like to be, confirming the axiom; conversely, an experience that does not exist is inconceivable [ii].

IIT's *axioms* 1 to 5 are meant to capture the *essential* properties of consciousness—those that are irrefutably true of every conceivable experience (Fig. 1A). Because they represent the essential properties of existence, the axioms of IIT are truly 'axiomatic:' unlike the axioms of mathematics, they are meant to be irrefutably true and not merely convenient starting points. The five axioms, just like the $0^{th}$ axiom, can be verified on oneself through introspection (provided their meaning is properly understood).

*(1)* *Intrinsicality*: every experience is *intrinsic*—it exists *for itself*. In other words, 'my' experience is for me ('subjectively,' without implying a separate entity that does the experiencing, even less as a self in a conceptual or autobiographical sense). Thus, when I experience a scene, the scene is for me, experienced from within, from my intrinsic perspective. This property is true of every conceivable experience: if I conceive of an experience that would be for someone else, it would then be theirs, confirming the axiom; conversely, an experience that is for no one is inconceivable.

*(2)* *Information*: every experience is *specific*—it is *this one*. Thus, when I see my bedroom, the experience I am having is this one, as opposed to another one. Again, this property is true of every conceivable experience: if I conceive of an experience that would be another one, it would then be that one, confirming the axiom; conversely, an experience that is not this one or any other one, but generic, is inconceivable. Because an experience is always this specific one, it implicitly differs from a large repertoire of other possible experiences, an attribute called *phenomenal differentiation*.

(3) *Integration*: every experience is *unitary*—it is *a whole*, irreducible to separate experiences. Thus, in front of me I see both the left and the right side of my bedroom, and what I see cannot be reduced to a left side and a right side that are experienced separately. Integration is true of every conceivable experience: if I conceive of an experience containing seemingly independent parts, like a 'flash' and a 'bang,' the experience would then be a 'flashbang,' which is a whole, confirming the axiom; conversely, an experience such that the left side would be experienced by me and the right side by someone else is inconceivable.

(4) *Exclusion*: every experience is *definite*—it is *this whole*, containing all it contains, neither less nor more. For example, my visual experience has a border: it includes all the visual field—its left and right side. It excludes my experiencing less—say, the left side only but not the right side—and my experiencing more—say, a periphery that extends to the back of my head. Exclusion is also true of every conceivable experience: if I conceive of an experience containing less or more, it would then be that whole, confirming the axiom; conversely, an experience that is not all it is, but indefinite, is inconceivable.

(5) *Composition*: every experience is *structured*—it is *the way it is*, being composed of distinctions and the relations that bind them, yielding a phenomenal structure that feels the way it feels. For instance, I can distinguish a body, a hand, and a book; the hand is attached to the body and lying on the book. Again, composition is true of every conceivable experience: if I conceive of an experience that might be some other way, it would then be structured that way, confirming the axiom. Conversely, an experience that is not this way or that way, but no way, is inconceivable [iii, iv].

By the same token, any other property of experience is not essential but *accidental*, in the sense that it may or may not be present. For example, I can certainly conceive of experiences that do not contain people, or sounds, or pains. It may be a bit harder, but I can also conceive (and achieve) experiences that do not contain a notion of self, that are not suffused by emotions, and even of experiences that to not feel extended in space or flowing in time.

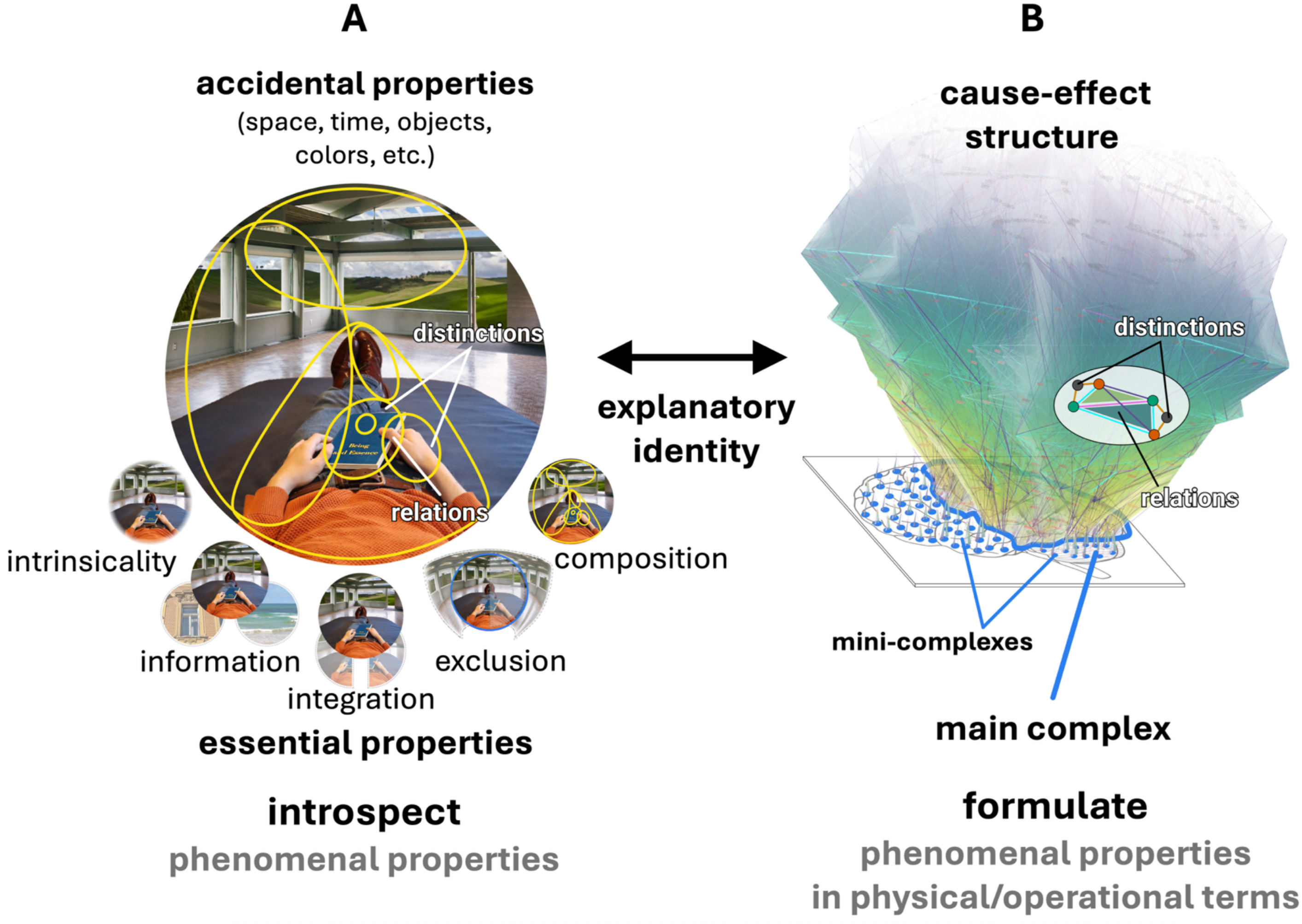

*Figure 1. IIT's explanatory identity.* (A) IIT starts from experience as intrinsic existence. Through introspection, one can identify five essential properties that are true of every conceivable experience—the axioms of phenomenal existence (A, bottom). (B) IIT then formulates the five axiomatic properties of experience in physical terms (understood operationally), yielding the postulates of physical existence. By applying the postulates to a system such as the brain, one should be able to identify a large substrate of units that satisfies those essential properties in terms of cause-effect power—the 'main complex'—corresponding to the neural substrate of consciousness. Many 'mini-complexes' outside the main complex, which would not be conscious in any meaningful way, would correspond to 'ontological dust' (B, bottom). The postulates are then applied to every subset of units of the complex to 'unfold' its cause–effect structure, composed of causal distinctions (causes and effects specified by each subset) and causal relations (the way distinctions overlap over their causes and/or effects) (B, top). According to IIT, the properties of the cause–effect structure account for all accidental properties of a given experience (such as space, time, objects, and colors) without any additional ingredients (A, top).

It is also important to recognize that the definition of consciousness as phenomenal existence has nothing to say about whether the experience occurs when we are awake and interacting with the environment, or asleep and dreaming. As long as there is experience, something genuinely exists.

**IIT's postulates: formulating the essential properties of phenomenal existence in physical (operational) terms**

To obtain an objective account of experience, the next step is to formulate phenomenal existence and its five axiomatic properties in terms of physical existence (Fig. 2 and 3), where 'physical' is understood as something that can be shown operationally through observations and manipulations, i.e. 'objectively.' *Physical existence*, IIT's $0^{th}$ postulate, is defined operationally as *cause–effect power*—the ability to 'take and make a difference.'

Axioms 1 to 5 can then be expressed in physical terms as properties of cause-effect power of a substrate, where a substrate is simply something that can take and make a difference—for example, a brain and its constituting neurons, represented in 2D in Fig. 1B. Paralleling the five axioms of phenomenal existence, IIT's *postulates* of physical existence are as follows:

*(1)* *Intrinsicality*: the cause–effect power of a substrate of consciousness must be *intrinsic*: it must take and make a difference *within itself*.

*(2)* *Information*: its cause–effect power must be *specific*: it must take and make a difference in *this state* and select *this cause–effect state*.

*(3)* *Integration*: its cause–effect power must be *unitary*: it must specify its cause–effect state as *a whole set* of units, irreducible to separate subsets.

*(4)* *Exclusion:* its cause–effect power must be *definite*: it must specify its cause–effect state as *this whole set* of units, all of them, neither less nor more.

*(5)* *Composition*: its cause–effect power must be *structured*: subsets of units must specify cause–effects over subsets of units (*distinctions*) that can overlap with one another (*relations*), yielding a *cause–effect structure* that is *the way it is* [v].

In formulating the postulates, IIT relies on several assumptions and principles aimed at providing a self-consistent 'objective' ontology and formulating its postulates [vi]. *Realism*—'there is a real world that exists (and persists) beyond my experience'—provides a good explanation for the many regularities of my experience (when I close and open my eyes, the room is always there). Its opposite—solipsism—may not be disprovable, but it explains nothing.

The *principle of being*—'to be is to have cause-effect power'—resembles the ancient definition of being known as the Eleatic principle. It defines objective or physical existence as 'taking and making a difference,' corresponding to *methodological physicalism*: This implies a 'substrate' of units that can be observed and manipulated (at least in principle), whose cause-effect power provides a good explanation for regularities in 'real world.' A substrate's cause-effect power is simply its transition probability matrix (TPM)—doing this produces that, with probability above chance—with no need for 'intrinsic properties,' such as mass and charge, or 'laws' governing their behavior [vii]. IIT's $0^{th}$ postulate is simply physicalism applied to the substrate of consciousness [viii].

The *principle of becoming*—'powers become what powers do'—governs how powers evolve as the substrate updates its state. It provides a good explanation for why powers are the way they are—that is to say, why the TPM is the way it is and how it changes [ix].

Two further ontological principles—minimal and maximal existence—are critical for formulating IIT's postulates. These principles are invoked to decide what exists, based on the definition of existence as cause-effect power, whenever there is a choice among multiple possibilities. The *principle of minimal existence*—'nothing exists more than the least it exists'—provides a sufficient reason for why, for example, one should take the minimum information partition when evaluating to what extent a system exists as one system (a system cannot be more integrated than across its weakest link). The *principle of maximal existence*—'what exists is what exists the most'—provides a sufficient reason for why, for example, among candidate system competing for existence over the same substrate, the one that actually exists should be the one that lays the greatest claim to existence (it is maximally integrated) [x].

Two further assumptions, finitism ad atomism, are required if one is to define minima and maxima. *Finitism*—'the substrate of cause-effect power is finite'— assumes that a complete explanation of physical existence should be based on a 'universal substrate' constituted of a finite number of units. *Atomism*—'the substrate of cause-effect power has a finest grain'—assumes that a complete explanation of physical existence should be based on the 'smallest' units that can take and make a difference, updating their state in discrete steps.

*Identifying complexes*

Much of the development of IIT has focused on formulating the postulates of physical existence in mathematical terms, to capture the essential properties of phenomenal existence faithfully, precisely, and uniquely. This is because, since every experience is unambiguously the way it is, its physical correspondent must also be unambiguous.

To satisfy intrinsicality, a candidate complex must have cause-effect power upon itself. To satisfy information, it must be in a specific state (its 'current' or 'actual' state) and select a specific cause state and a specific effect state upon itself. Based on the principle of maximal existence, its selected cause and effect state is the one with maximal *intrinsic information* ($ii_s$), a measure that satisfies existence, intrinsicality, and information uniquely (Barbosa et al., 2020). Intrinsic information requires both intrinsic differentiation (the ability to provide itself with a repertoire of cause and effect states of non-zero probability) and intrinsic specification (the ability to specify a state by increasing its probability) [xi].

To satisfy integration, a candidate complex must be irreducible. This is assessed by system *integrated information* ($\varphi_s$), the intrinsic information across the system's minimum information partition, in accordance with the principle of minimal existence. Finally, to satisfy exclusion, a complex must include a definite set of units, neither less nor more. According to the principle of maximal existence, this is the set of units for which integrated information is maximal ($\varphi_s^*$). This set of units is then identified as the first-maximal or *main complex* over the substrate (Marshall et al., 2023) [xii]. The remaining units in the substrate are then assigned recursively to non-overlapping complexes (second-maximal complex, third-maximal complex, and so on) until all units are exhausted. A substrate can thus be said to 'condense' into a number of non-overlapping complexes, or 'intrinsic entities,' each of which satisfies the requirements for phenomenal existence in physical terms. Some may be very large, and a multitude of others

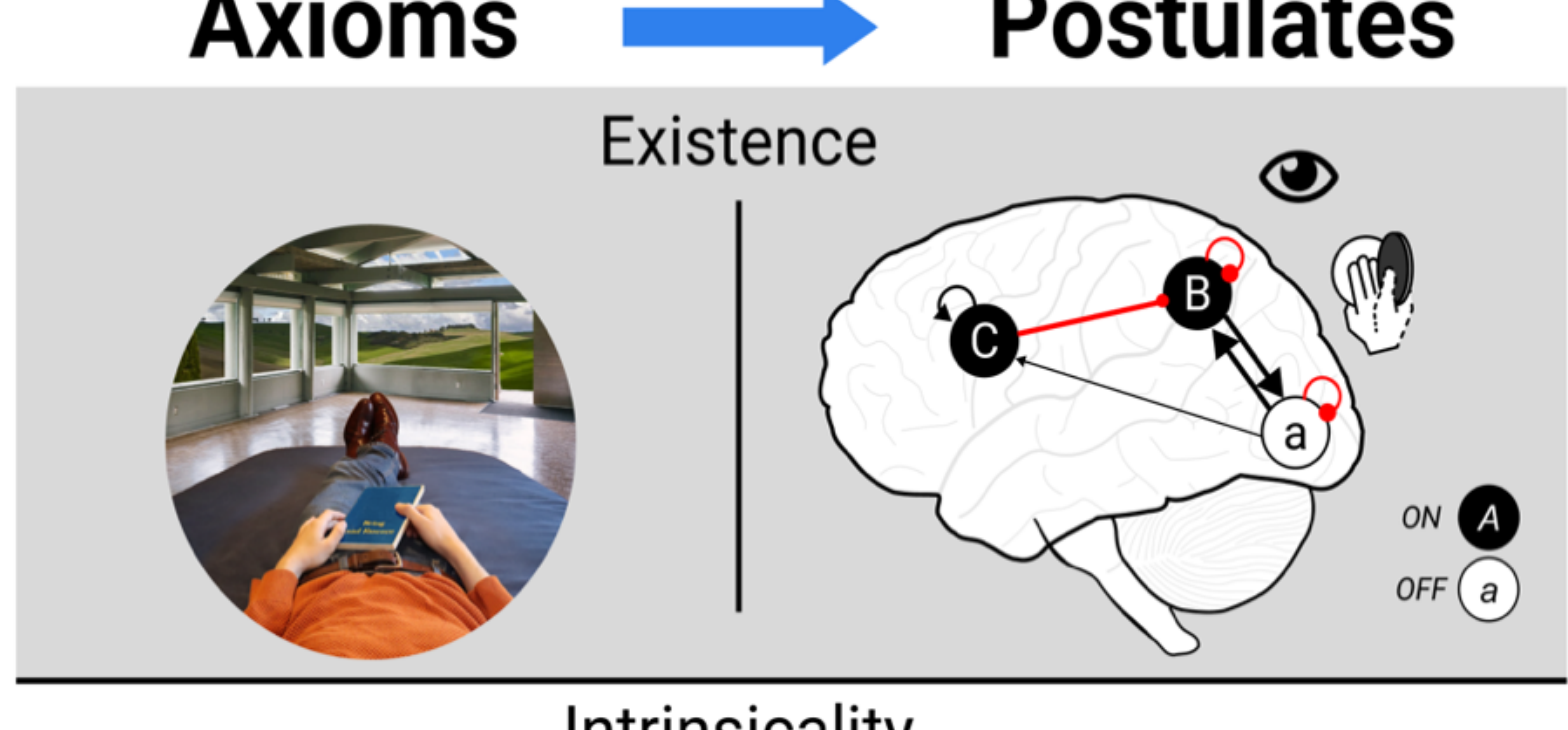

Experience exists. In physical terms, this corresponds to the existence of a substrate (a set of units in a state, e.g., the brain, or here ***aBC***, where lowercase indicates OFF and uppercase indicates ON) with the ability to 'take and make a difference' from the perspective of a conscious observer and manipulator (indicated by the 'eye' and 'hand').

Intrinsicality

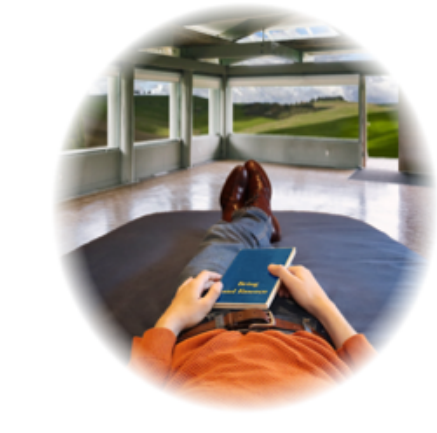

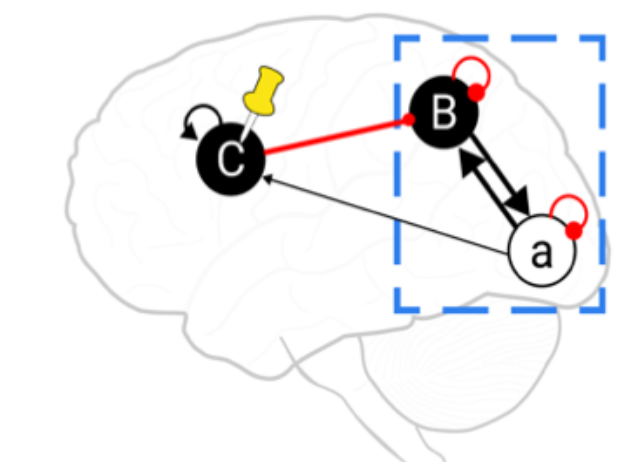

Just as every experience is *intrinsic*—it exists *for itself*, its substrate must be *intrinsic*—it must take and make a difference *within itself*. The dotted blue line indicates a candidate substrate (***aB***) whose cause-effect power can be analyzed within itself.

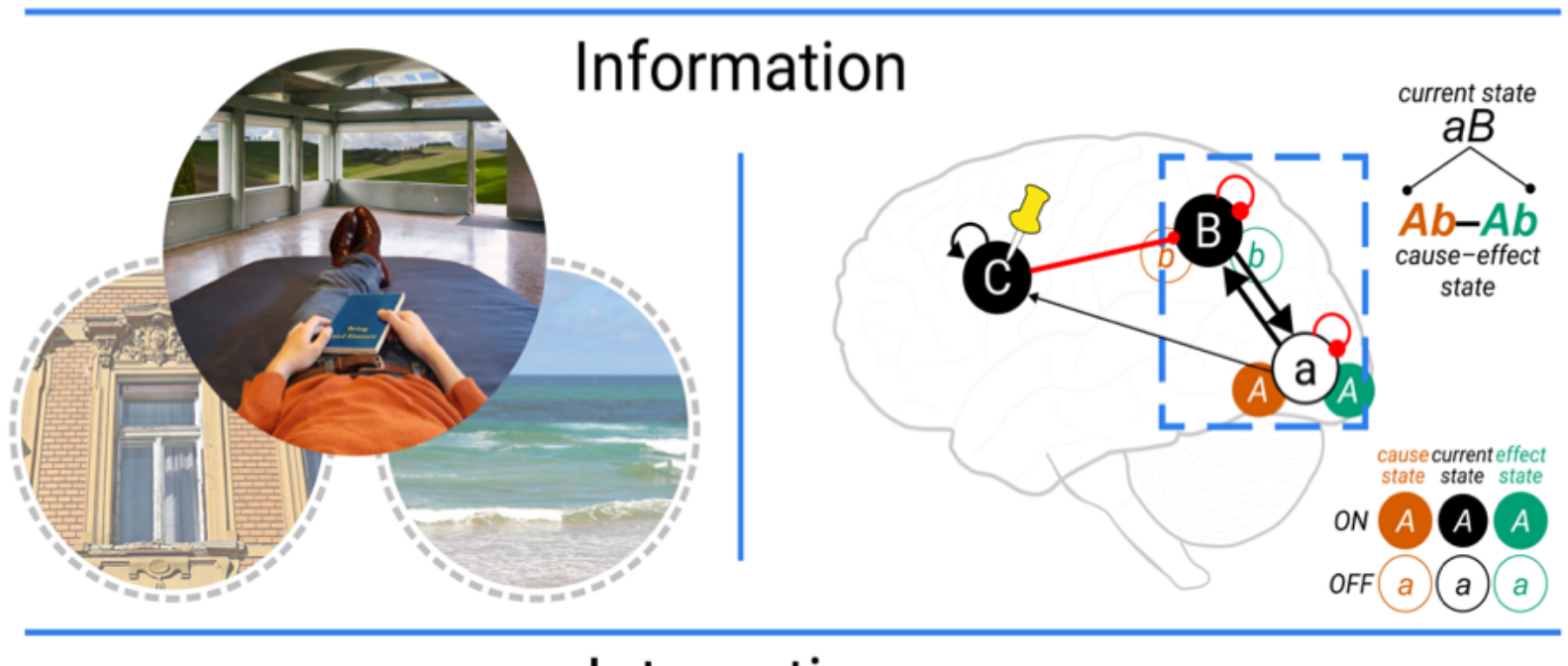

Just as every experience is *specific*—it is *this one*—its substrate must be in a *specific* state, selecting *this cause-effect state*, the state that maximizes intrinsic information ($ii_s$), here ***Ab–Ab*** (red for cause and green for effect). And just as every experience, by being specific, differs from countless other experiences, so does each cause–effect state specified by different current states (e.g., ***ab*** or ***AB***).

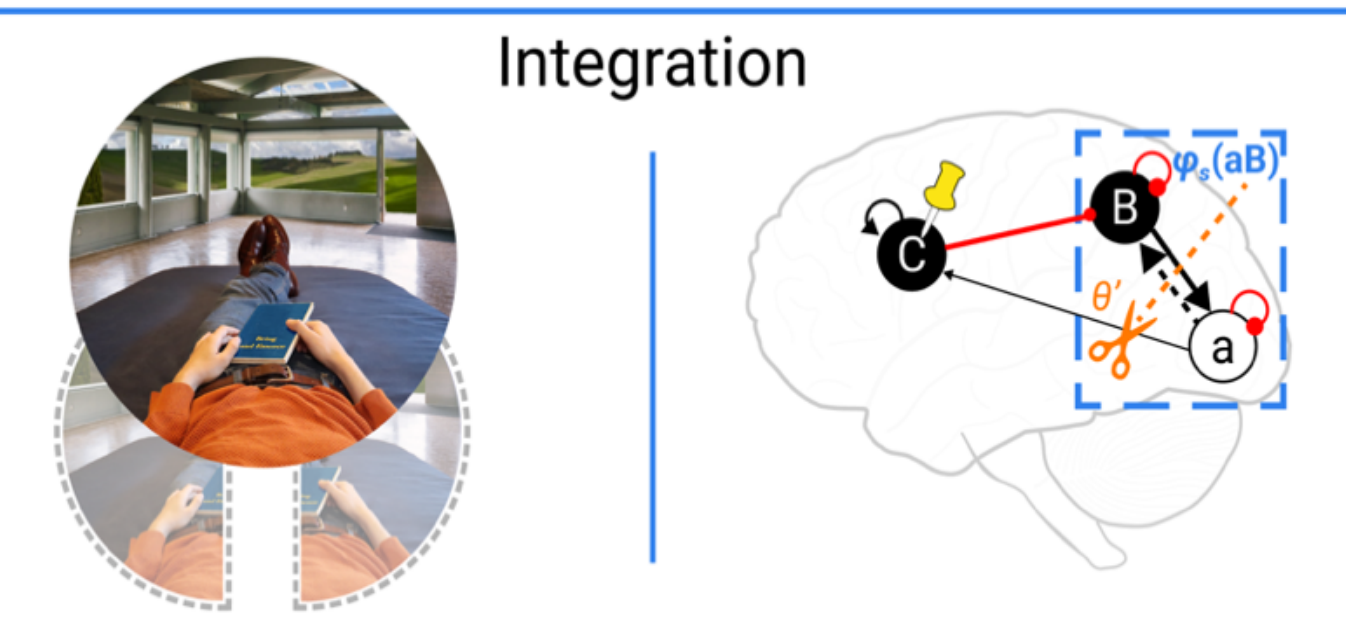

Just as every experience is *unitary*—it is *a whole*, irreducible to its parts—the cause–effect power of its substrate must be *unitary*—it must specify its cause–effect state as *a whole set* of units (say, ***aB***) that is irreducible to separate subsets (e.g., to ***a*** and ***B***, when the substrate is partitioned unidirectionally by cutting the connection from ***a*** to ***B***, as indicated by the dotted line). Irreducibility is measured by integrated information ($\varphi_s$).

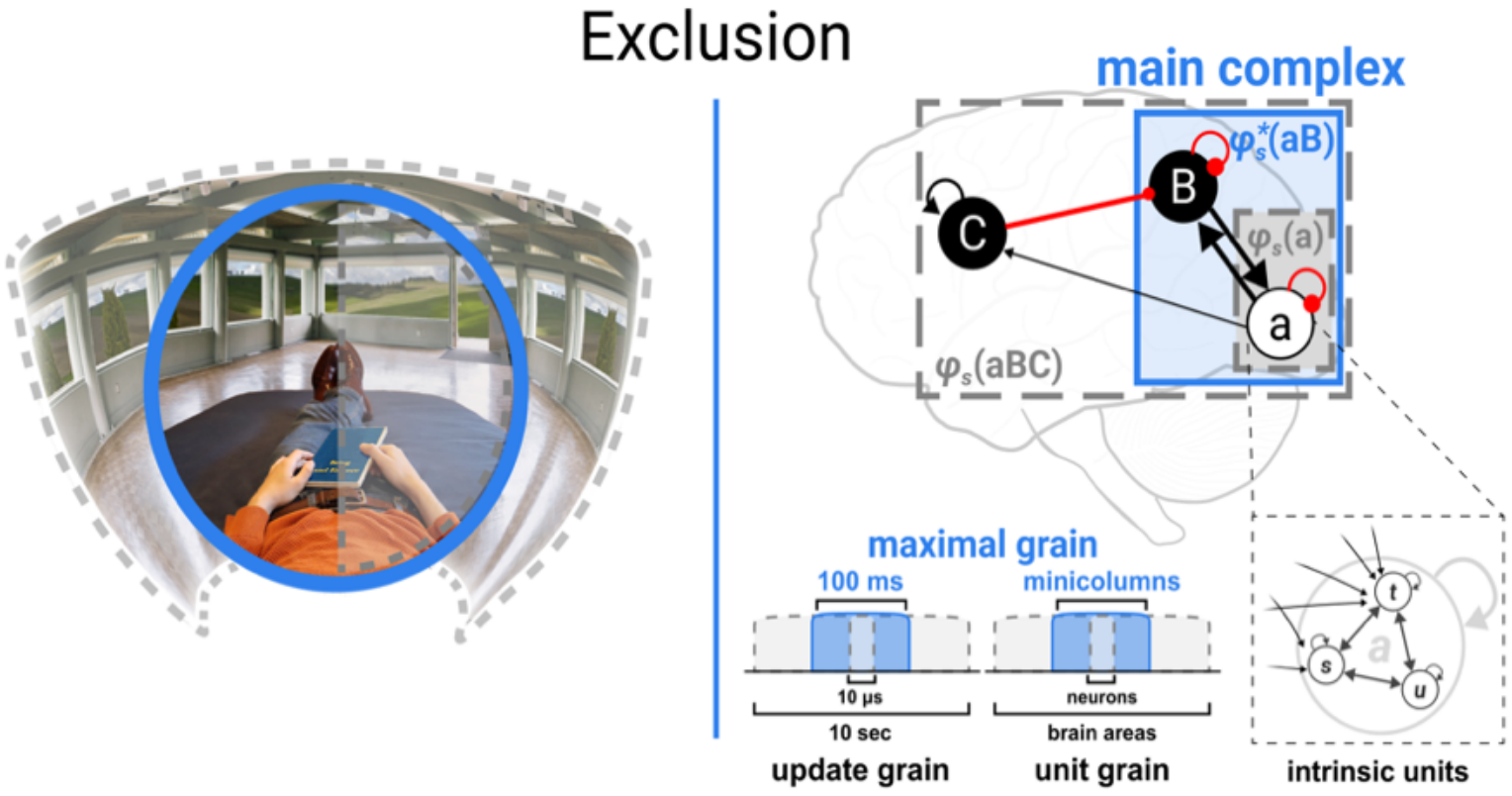

Just as every experience is *definite*—it is *this whole*—the cause-effect power of its substrate must be *definite*—corresponding to the set of *intrinsic units* that is maximally irreducible (called the *main complex*). The main complex (here ***aB***, indicated by the blue outline) is the one for which integrated information is maximal ($\varphi_s^*$) and whose intrinsic units are maximally irreducible within, thus excluding all overlapping sets (e.g., ***a*** and ***aBC***) and all overlapping unit and update a grains (e.g. micro units ***s***, ***t***, and ***u*** with shorter updates).

*Figure 2: Identifying complexes by applying the postulates of existence, intrinsicality, information, integration, and exclusion.*

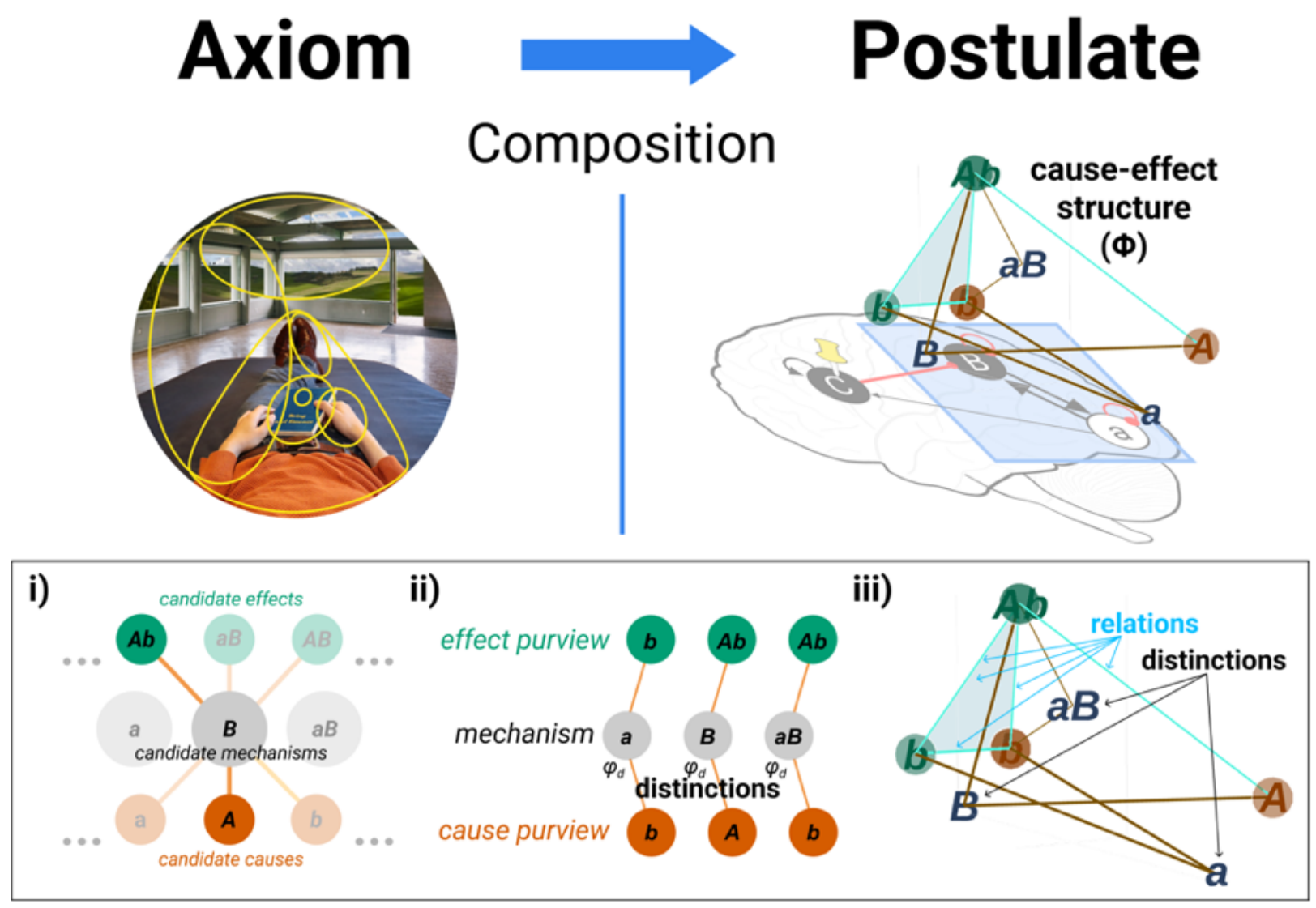

Just as every experience is *structured*—it is *the way it is*, being composed of phenomenal distinctions and the relations that bind them—the cause–effect power of its substrate must be *structured* by causal distinctions and relations, yielding a cause–effect structure that is *the way it is*. The cause–effect structure is unfolded from its substrate (the main complex identified based on the previous postulates) by considering subsets of its units and: **(i)** assessing candidate distinctions, where mechanisms (***a***, ***B***, ***aB***, in black) specify causes and effects (in red and green, respectively); **(ii)** assessing whether they are maximally irreducible, as measured by $\boldsymbol{\varphi_d}$, and congruent with the complex's cause-effect (*causal distinctions*); **(iii)** if so, assessing overlaps among causes and/or effects, whose irreducibility is expressed by $\boldsymbol{\varphi_r}$ (*causal relations*). Distinctions and relations compose the cause–effect structure specified by the complex in its current state. The sum of their $\boldsymbol{\varphi}$ values corresponds to the complex's *structure integrated information* ($\boldsymbol{\Phi}$).

*Figure 3: Unfolding a complex's cause-effect structure by applying the composition postulate.*

minuscule (Fig. 1B). For the intrinsic perspective of every complex, the substrate units that do not belong to it serve as *background conditions*.

In general, a system of many units will have a larger repertoire of possible states (greater maximal differentiation), which allows for greater values of intrinsic information. However, the ability to specify causes and effects within the system tends to decrease with more units because cause and effect information is spread over additional states. This implies that the system's units must be able to interact very effectively and selectively(Barbosa et al., 2020; Marshall et al., 2023). Moreover, a larger system can only hold together well if its units are appropriately interconnected, say as a dense lattice, rather than organized in a modular manner. Otherwise, 'fault lines' will develop, with the consequence that values of $\varphi_s$ would be higher for smaller subsets. These would then beat the larger system in the 'competition for existence' (Marshall et al., 2023).

In the brain, much of the cerebral cortex, primarily its posterior-central portions, appears well suited to satisfy these requirements. Pyramidal neurons, particularly in supragranular layers, are highly specialized and grouped within densely connected minicolumns. Furthermore, if the background conditions are kept nearly constant, groups of neurons appear capable of interacting in a highly effective manner. The dense, divergent-convergent hierarchical lattice of connections among specialized units, each implementing different input–output functions, but partially overlapping in their inputs (receptive field) and outputs (projective fields), is especially well suited to constituting a large complex, minimizing fault lines (Albantakis et al., 2023).

In general, sets of units with values of $\varphi_s$ higher than *most* overlapping sets—that is, *relative* or *extrinsic* maxima of integrated information—are likely to capture relevant levels of substrate organization by 'carving nature at its joints.' For example, the human body, as a paradigmatic organism that is highly integrated within and clearly demarcated without, likely constitutes a relative maximum and can be considered as a well-defined *extrinsic entity*. Similar considerations apply to organs such as the heart, the liver, and the brain taken as a whole. However, as discussed in a later section, there is a critical difference between these relative or extrinsic entities and the *absolute* maximal complex over a substrate—an *intrinsic entity*: only the latter constitutes a substrate of consciousness and contributes to the way an experience feels—all other entities do not exist from the intrinsic perspective. The divide between extrinsic and intrinsic entities—between what exists for itself and what only exists for something else—can be considered as the *great divide of being* (Tononi et al., 2022).

*Identifying a complex's intrinsic units*

Another consequence of IIT's postulates is that the units that constitute a complex—its *intrinsic units*—may be macro rather than micro units—that is, they may be constituted of multiple micro units and update their state over multiple micro updates (Marshall et al., 2024). Like the complex itself, its units must comply with IIT's postulates of physical existence, adjusted to their status as units (rather than complexes). Thus, units must have cause-effect power upon themselves (intrinsicality), in a way that is specific (information) and unitary (integration). For example, integration ensures that a complex is not built out of macro units that do not themselves exist because they are reducible to their parts (otherwise one could build something out of nothing). On the other hand, while units must be definite (exclusion), exclusion only requires that units be maximally irreducible 'within' (they must have greater unit $\varphi_s$ than any combination of their constituents). This is because units do not exist as complexes themselves, but as constituents of a complex. Moreover, the set of non-overlapping intrinsic units must be such that, taken together, they maximize the complex's irreducibility $\varphi_s$.

Macroing also applies to macro states defined over different update grains. For example, many different sequences of states may be macroed into one macro state, and the remaining sequences into the alternative macro state. Over multiple micro updates there can thus be a large number of possible mappings of sequences of micro states into two macro states. Again, the mapping that complies with IIT's postulates and principles is the one that maximizes the complex's $\varphi_s$. From the perspective of the complex, intrinsic units only exist in one of two alternative macro states and have no internal structure of their own.

By considering all possible grains for small systems of binary units, it can be shown that a system can indeed have greater $\varphi_s$ if its constituent micro units are apportioned into macro units (Hoel et al., 2013) (Marshall et al., 2024). This can be the case if a macro grain increases intrinsic information [xiii]. Moreover, a system of macro units can have greater cause-effect power than the corresponding micro units if integration is higher at the macro level. The determination of a complex's intrinsic units must also take into consideration the recursive macroing of micro units into macro units. It can be conjectured that macro units built upon a hierarchy of finer units (also called 'meso' units) may be critical for allowing large systems to exist as maxima of intrinsic, irreducible cause-effect power. Hierarchies of this sort appear to be a common feature of biological systems.

In general, macro grains with $\varphi$ values higher than *most* finer or coarser grains are also likely to capture relevant levels of substrate organization, even though they may not be relevant for consciousness. In the brain, for example, macro grains might correspond to proteins, ion channels, organelles, synaptic vesicles, synapses, neurons, groups of tightly interconnected neurons, and so on. Similarly, macro states might correspond to relevant time constants, such as those for ion channel openings, membrane time constants, the time constants of AMPA or NMDA receptors, those favorable to inter-areal interactions, response times, and so on. Such 'extrinsic units' and 'extrinsic states,' well-suited to manipulations and observations by neuroscientists, are important for understanding how the system works. However, according to IIT, there is a critical difference between these relative maximal grains and the absolute maximal grain whose intrinsic units and states maximize $\varphi_s$ within and without: only the latter constitutes the substrate of consciousness and contributes to the way the experience feels—all other levels of organization do not exist from the intrinsic perspective.

In summary, by applying postulates 1 to 4, IIT concludes that the substrate of consciousness must be a large set of units that: (1) has cause-effect power over itself (intrinsicality); (2) has a specific state with a specific cause and effect (information); (3) is irreducible (integration); and (4) maximally so (exclusion). Maximal irreducibility, measured by integrated information ($\varphi_s^*$), requires adequate anatomical and physiological properties, such as a dense lattice of connections among specialized macro units capable of effective causal interactions.

*Unfolding a complex's cause-effect structures*

Once a complex and its intrinsic units have been identified (Fig. 2), the final postulate, composition (Fig. 3), requires that we fully *unfold* its cause-effect structure by considering the cause-effect power of all its subsets of units—the causal distinctions they specify (their causes and effects within the complex) as well as the causal relations among them (the overlaps among causes and/or effects) (Albantakis et al., 2023). The cause-effect structure unfolded from a complex is represented in 3D in Fig. 1B.

The convention of illustrating intrinsic entities as 3D $\Phi$-structures unfolded from a 2D substrate is meant to convey graphically a central notion of IIT: *what genuinely exists, in physical terms, is not a substrate as such—a bare substrate—but the substrate unfolded into the $\Phi$-structure it specifies—an intrinsic entity.* Indeed, a useful way of thinking of a complex as a substrate is simply as the *folded* (implicit, packed, compressed, or reduced) summary of what actually exists. In other words, a substrate merely represents a set of units that can be observed, perturbed, pinned, and partitioned, whose cause-effect power is tabulated in the corresponding TPM. To determine what actually exists, a complex must be *unfolded* (explicated, unpacked, uncompressed, or expanded) according to the postulates of composition, to reveal its full cause-effect power [xiv].

Like the complex itself and its intrinsic units, causal distinctions and relations are assessed by considering IIT's postulates (save for composition). A subset in its current state (a mechanism) that links a cause and an effect over subsets of complex's units (its cause and effect purview, respectively) forms a candidate distinction that satisfies existence. As for the system as a whole, the cause-effect state specified over the purviews is the one with maximal intrinsic information (*ii*), satisfying intrinsicality and information. A candidate distinction can only exist for the complex if it is irreducible, satisfying integration (measured by distinction integrated information $\varphi_d$), and maximally so, satisfying exclusion ($\varphi_d^*$). Note that the distinctions that actually exist for the complex are only those whose cause–effect state is congruent with the cause–effect state of the complex as a whole [xv]. The upper bound on the number of distinctions specified by a complex of n binary units is 2^n -1 (A. Zaeemzadeh & G. Tononi, 2024).

Finally, overlaps among causes and effects of one or more distinctions capture additional irreducibility of the complex's cause-effect power. The number of causes and/or effects bound by a relation are its *faces*. Being composed of distinctions, relations satisfy IIT's postulates and are necessarily congruent. Their irreducibility, measured by $\varphi_r$, is assessed by 'unbinding' distinctions from their joint purviews, taking into account all faces of the relation. An upper bound on the number of relations is exceedingly large, i.e. 2^(2^n -1) -1 (A. Zaeemzadeh & G. Tononi, 2024).

Together, these distinctions and relations compose the *cause–effect structure* of the complex in its current state, also called a *$\Phi$-structure*. The sum of its distinction and relation integrated information amounts to the structure integrated information of the complex ($\Phi$, or *big PHI*). In short*, the cause-effect structure captures all that a complex specifies about itself, in causal terms, through its various subsets*.

**IIT's explanatory identity: an experience as a cause-effect structure**

IIT's *explanatory identity* claims that the cause-effect structure unfolded from a complex in its current state should account for *all* the properties of the experience it supports, *with no additional ingredients* [xvi].

The *essential* properties of experience—that every experience is intrinsic, specific, unitary, definite, and structured—are accounted for explicitly through the requirements imposed on the substrate of consciousness by the postulates of IIT. By the explanatory identity, the *accidental* properties that make an experience feel the way it feels—structured in that particular way—must also be accounted for by corresponding properties of the $\Phi$-structure. Thus, the feeling of spatial extendedness, of temporal flow, of objects, of colors and sounds, of thoughts and emotions, and so on, should find correspondence in the way the $\Phi$-structure is organized. In short, *quality is structure*.

The quantity of consciousness—the degree of intrinsic existence—would also correspond to a property of the $\Phi$-structure—namely in its $\Phi$ value—the sum of the $\varphi$ values of all its component distinctions and relations. Whenever consciousness fades, then, its substrate and associated $\Phi$-structure should disintegrate;

whenever it returns, it should reintegrate. Moreover, the degree to which a content exists within an experience should be accounted for by the degree to which a *sub-structure* (also called *Φ-fold*) exists within the *Φ*-structure.

Properties summarizing overall features of the experience—such as its subdivision into modalities, its vividness, and so on—should also be accounted for by corresponding features of the *Φ*-structure. Finally, the similarity/dissimilarity of contents within an experience should be accounted for by the similarity/dissimilarity of their respective *Φ*-folds, and similarities/dissimilarities between experiences by similarities/dissimilarities between the corresponding *Φ*-structures.

The explanatory identity of IIT is an exacting but powerful feature: with just five postulates of physical existence and a substrate in a state, IIT conjectures that it should in principle be possible to fully characterize any experience as a cause–effect structure. For IIT, in other words, *the actual is the potential*, in the sense that what something is (the actual) is given by its powers (the potential), properly unfolded.

Three points should be emphasized. First, the identity between phenomenal existence and physical existence is not between two separate substances or domains of existence. Phenomenal existence exists intrinsically—for itself, i.e. absolutely. Physical existence is defined operationally as cause-effect power—taking and making a difference—as judged by us as conscious beings, i.e. relatively. The identity is explanatory because it tries to account for one sense of existence—phenomenal existence, which is intrinsic or subjective—through another sense of existence—physical existence, which is defined operationally or objectively.

Second, the properties of *Φ*-structures offer the prospect of decomposing the properties of experience in a way that is much finer than what can be done by introspection alone.

Finally, the same kind of explanation, where existence is defined operationally as cause-effect power, can be applied to any natural phenomenon. In principle, one can imagine a unified account of nature, where intrinsic existence (consciousness) and extrinsic existence (bodies, rocks, and stars) can be analyzed using the same tools—the postulates of physical existence—formulated based on the properties of intrinsic existence, as revealed by experience itself.

## Empirical validation: explanations and predictions

Characterizing the essential properties of experience—those that are true of every conceivable experience—leads to testable predictions about the necessary and sufficient conditions for the substrate of consciousness. Moreover, accidental properties of experience—that it usually feels extended in space, flowing in time, containing objects and narrow qualia—should be accounted for by the way that substrate is organized.

Because consciousness is subjective—it exists from the intrinsic perspective of an experiencing subject—the empirical evidence for IIT must ultimately come from us—adult humans who can report our experiences. IIT's empirical validation program began long ago by considering basic facts about when we are conscious or not. Why is it that, for us to be conscious, the corticothalamic system must be intact, but not the cerebellum or the spinal cord? Why is it that, during slow-wave sleep, we can lose consciousness even though the corticothalamic system remains active? The program then expanded to address the quality of consciousness. Why is it that what we see around us or feel on our body is experienced as extended—painted or etched on the canvas of space? And why is it that what we hear, sentences or melodies, feels flowing—played on the track of time?

**The presence of consciousness: the main complex**

By requiring that the neural substrate of consciousness satisfies the postulates of intrinsicality, information, and integration, IIT provides a principled explanation for several well-established facts. For example, given the postulates, we can understand why certain parts of the cerebral cortex can support consciousness, and thereby constitute the *main complex*, whereas the cerebellum does not (Tononi et al., 2016). According to IIT, this is because their connectivity, especially that of posterior-central cortex, is ideally suited to support high integrated information $\varphi_s$. In topographically organized cortical areas, neurons are linked by a grid-like horizontal connectivity, augmented by converging–diverging vertical connectivity along sensory hierarchies. By contrast, cerebellar micro-zones are largely independent of one another and are organized in a feedforward manner, hence they cannot constitute a large complex. These differences in organization can explain why widespread lesions of posterior-central cortex can affect consciousness directly, whereas widespread lesions of the cerebellum, which has four times more neurons, do not.

However, having the right anatomy is not sufficient. The loss of consciousness during periods of slow-wave sleep can be explained, based on IIT's postulates, by the breakdown of integrated information (Tononi et al., 2024). Due to changes in neuromodulation in slow-wave sleep, neurons become bistable, in the sense that activation is rapidly followed by a stereotypical OFF-period, lasting tens to hundreds of milliseconds. This underlies a collapse of intrinsic information, since most inputs lead to the same output—namely an OFF-period—and consequently to a collapse of integrated information. Experiments in subjects with epilepsy have confirmed that, while electrical stimulation of the cortex during wakefulness triggers a chain of phase-locked activations, during slow-wave sleep the same input triggers a stereotyped slow wave associated with a cortical OFF-period (Pigorini et al., 2015). Cortical activity then resumes, but the phase-locking to the stimulus is lost, indicative of a breakdown of causal interactions. The loss of consciousness during some seizures may have a similar explanation: when neural activity is paroxysmal, most inputs make no difference to the output.

The requirement that the substrate of consciousness must have a large repertoire of states yet be highly integrated within has led to the development of methods for testing IIT through the brain's responses to perturbations (Massimini et al., 2005). This can be done by perturbing the cerebral cortex using transcranial magnetic stimulation (TMS) to engage near-deterministic interactions among distributed groups of cortical neurons and estimating the repertoire of available states through the complexity of the resulting responses. An index reflecting the complexity of these responses, the perturbational complexity index (PCI), can thus serve as a crude proxy of the brain's capacity for integrated information (Casali et al., 2013). In a series of experiments, it was shown that PCI is invariably high when subjects are conscious, whether awake or dreaming (including during vivid dreams under ketamine anesthesia), and low when consciousness is lost in dreamless sleep or dreamless anesthesia (Casarotto et al., 2016).

IIT's exclusion postulate adds a further requirement: the substrate of consciousness—the main complex—must have a definite border and grain, corresponding to maximal integrated information. As already mentioned, the cerebral cortex, unlike the cerebellum, seems well-suited to ensuring high values of integrated information. Broadly speaking, the entire cortex is highly 'integrated'—it is a paradigmatic example of a richly interconnected network. On the other hand, given the exclusion postulate, the main complex should not just be integrated, but maximally so. Accordingly, IIT conjectures that the maximum of integrated information may correspond primarily to posterior-central cortical areas, owing to the dense, lattice-like connectivity within and among groups of neurons organized topographically and hierarchically. In contrast, the local connectivity within many prefrontal areas appears to have a modular organization (augmented by diffuse projections) (Watakabe et al., 2023). Therefore, these areas would be excluded from the main complex, despite an equal or larger number connections, because 'fault lines' among modules reduce integrated information. While these predictions of IIT fit with evidence from lesion, stimulation, and recording studies (Boly et al., 2017; Koch et al., 2016), they remain controversial and are subject of ongoing empirical tests (Cogitate et al., 2025).

Another controversial prediction of IIT is that the localized activation of early sensory areas within the main complex should support the experience of simple stimuli (visual, auditory, and somatosensory) without requiring widespread activations of other areas within the main complex, and without the need for global 'broadcasting,' 'ignition,', or higher order 'monitoring.'

Prefrontal areas are involved in many long-range pathways mediating complex neuronal interactions within and outside the cortex. Circuits involving prefrontal modules are ideally suited to support *processing loops* that mediate cognitive functions such as attention, reasoning, reflection, and memory. These and longer processing loops through the cerebellum and the basal ganglia would affect the main complex powerfully but indirectly, without being included in it (this is because minimal partitions across directed cycles originating and ending in the main complex typically yield a much lower value of $\varphi_s$ compared to minimal partitions across the complex's lattice). Similarly, sensory and motor pathways would remain outside the main complex despite their key role in triggering specific contents of experience and carrying out complex behaviors.

In general, according to IIT, any brain constituent that qualifies as a unit of the main complex must contribute directly to experience, whereas anything outside the main complex does not. Brain constituents that are essential to the cause-effect power of the main complex but are not part of it are considered as *background conditions* (Albantakis et al., 2023). These include, besides processing loops and input-output pathways, anything from local blood supply to arousal systems, which ensure the proper functioning of the main complex.

A further prediction of IIT is that even units within the relevant parts of posterior-central cortex may not contribute to experience if the causes and effects they specify are *incongruent* with the cause–effect state specified by the main complex as a whole. This may happen, for example, during dreaming sleep, when some neurons may respond to external stimuli in a way that does not fit with the current dreamt experience (Tononi et al., 2024). Or it may occur during binocular rivalry, when stimuli presented to one eye may affect neurons in posterior cortex in a way that is incongruent with the dominant percept.

Broadly defined, phenomenal *dissociations* provide further test cases for IIT. The splitting of consciousness after surgical resections of the corpus callosum is an obvious example, especially if the phenomenal split can be attributed primarily to posterior-central resections, as seems to be the case (Sauerwein & Lassonde, 1997). An intriguing prediction that follows from the exclusion postulate is that, if one could progressively reduce the efficacy of callosal transmission, there would be a moment at which a single consciousness would suddenly split into two after a minor change in the traffic of neural impulses across the callosum. A dissociation from a single main complex might also occur in functional blindness, when a patient are subjectively blind but may purposefully avoid obstacles, and in other dissociative disorders. An important open question is the extent to which the constitution of the main complex may be modulated by selective attention: can certain cortical areas 'drop out' due to reductions in excitability mediated by attentional mechanisms? What happens when one has an engaging conversation and pays no attention to driving? Does one still 'see' the road while failing to 'notice' it, does the road vanish from consciousness, or does consciousness split in two, a talking one and a driving one? (Sasai et al., 2016).

Finally, as discussed above, the exclusion postulate has implications for the grain and state of the units of the main complex—its *intrinsic units*. For example, assume that the grain of intrinsic units is that of, say, minicolumns, and the update grain on the order of, say, 30 milliseconds [xvii]. In that case, experience should only change if there is a change in the state of intrinsic units at that grain. Any other changes should affect the brain but not experience. More generally, any change in unit micro state that does not translate into a switch of their macro state will not affect experience. For instance, changes in the timing or rate of firing of an individual neuron may have measurable effects on the rest of the brain, but if they map onto the same intrinsic macro state, they will not change experience.

**The quality of experience: space is extended**

IIT's explanatory identity predicts that the quality of an experience—the way it feels—should be accounted for by the cause-effect structure unfolded from the main complex in its current state, with no additional ingredients. The first attempt to evaluate this prediction focused on the way space feels, namely that it feels *extended* (Haun & Tononi, 2019). There were two main reasons for this choice. One is that the feeling of spatial extendedness is pervasive. Our body feels extended (personal space), we can reach out in the space immediately around us (peri-personal space), the visual field reveals a wider horizon (distal space) and, most generally, we feel located in an environment (environmental space). 2D visual space is paradigmatic: whether we are awake or dreaming, all visual contents—colors, contours, and shapes—are 'painted' on the canvas of space—they are 'located.'

Another, critical reason is that the experience of space, especially visual space, is partially penetrable through introspection. To characterize what it takes for space to feel extended, we should first abstract away from objects or local qualities: the sky feels extended whether or not it contains stars, and whether it is blue or black. We can then deploy spatial attention to begin dissecting its structure. We can easily 'spotlight' or 'pick out' any portion of experienced space—small, large, or anything in between, here, there, or anywhere in between. For lack of a better term, we call these phenomenal distinctions *spots*. The way spots are related can also be introspected: every spot overlaps or 'points to' itself (reflexivity), is included by or includes other spots (inclusion), overlaps partially with other spots (connection), and fuses with a connected spot to form another spot (fusion). Other aspects of how space feels can be expressed in terms of these fundamental properties: for example, two spots may feel at a certain distance from each other, and that distance can be expressed as the smallest spots that is connected to both.

The next step is to envision which kind of substrate, when unfolded into a $\Phi$-structure according to the postulates of IIT, and with no additional ingredients, might account for the properties that compose the feeling of spatial extendedness. A natural guess is a grid-like substrate. In fact, it can be shown that the cause-effect structure unfolded from a grid is composed of distinctions that are related in a way that accords precisely with reflexivity, inclusion, connection, and fusion—just like spatial phenomenology. Such a $\Phi$-structure (or sub-structure) is called an *extension*.

Notably, most areas within posterior-central cortex are organized as grids, beginning with primary visual cortex. The prevalence of grid-like architectures in the portion of cortex that is the prime candidate for supporting consciousness fits well with the prevalence of spatial phenomenology. Furthermore, lesions of such areas result in the loss of experience of corresponding regions of the visual field; their electrical stimulation can evoke sensations that are arranged 'topographically;' and recordings show that perceived locations are systematically correlated with which position within such areas is activated (Haun & Tononi, 2019).

These well-known observations are often interpreted as indicative of a 'topographic mapping' of external stimuli, 'representing' their spatial position in the environment. The adaptive value of topographic mapping is beyond doubt, but mapping cannot explain why space feels the way it does, and why we experience space when we dream, disconnected from the environment (M. Grasso et al., 2021; Haun & Tononi, 2019). As will be discussed below, the feeling/meaning of space, like any other feeling/meaning, must be specified intrinsically. And indeed, patients with extensive lesions of grid-like visual cortex may not just become 'blind,' say in one half of the visual field. Instead, for them that half of space simply 'ceases to exist,' and they are typically not aware that anything is amiss.

An intriguing consequence of IIT is that changes in connectivity within the main complex should result in changes in experience even if they are not accompanied by changes in activity. An initial test of this prediction was performed by targeting the experience of space. During a training phase, two nearby spots were coflashed repeatedly to transiently enhance the connectivity between their cortical targets. After this training, the space between two distant, untrained spots was perceived as contracted, even though the cortical activity triggered by the two untrained spots was presumably unchanged (Song et al., 2017). Ongoing experiments are testing the prediction that patients with paracentral scotomas due to lesions of primary visual cortex should also experience space as contracted (Olcese et al., 2024). Further experiments could be envisioned to test this general prediction of IIT—that changes in connectivity should be associated with changes in experience even without changes in activity—including comparing experiential qualities before and after the refinement of connections during development.

Yet another consequence of IIT, often considered counterintuitive, is that a main complex that is largely *inactive* should support consciousness, as long as it is in a state of causal readiness, i.e. not *inactivated*. Under the assumption that the main complex in the human brain is largely constituted by 2D grids, one would also expect that, when inactive, it would specify a $\Phi$-structure that is fundamentally an extension. The corresponding experience should then be fundamentally spatial in nature. Intriguingly, states of 'pure presence' or 'naked awareness' achieved by long-term practitioners of certain meditation traditions are typically described as vivid experiences of vast, often luminous extendedness, devoid of thoughts, self, or phenomenal objects. Such states are hard to achieve, especially in laboratory settings. Even so, the results of an exploratory study indicate that, when trained meditators reach states of pure presence, their EEG shows a broadband decrease in power, most marked in the gamma band (compared to any other experiential states, including mind-wandering). A decrease of EEG power in the gamma range is suggestive or reduced neuronal firing (as long as muscle signals and other artifacts can be properly ruled out) and is compatible with IIT's prediction (Boly et al., 2024).

A few other observations may help better understand IIT's account of spatial experience. One is the remarkable philosophical 'blind spot' concerning the experience of space. The mind-body problem, or 'explanatory gap,' is regularly introduced with reference to the experience of color, sound, pain, and the like: it seems impossible for science to explain objectively why blue should subjectively feel like blue—in fact, why it should feel like anything at all. But hardly anybody has wondered why space should feel extended—a question that, given the pervasiveness of spatial experience, its partial accessibility to introspection, and the wealth of neural data, might offer a better chance of success. A likely reason for this blind spot may be that we are perpetually immersed in space, like fish in water. We thus take it for granted as the canvas upon which objects and colors are painted, rather than realizing that, without space, we could not experience either.

Another blind spot is the focus on the functional aspects of seeing, rather than on the structural ones—on what we can *do* with vision, rather than what vision *is like*. For most purposes, it is perfectly acceptable to treat the visual field, for example, as a coordinate system, convenient for communicating instructions. But there is nothing inherently spatial about a coordinate—a pair of numbers can only be assumed to convey a 'local sign' if placed in a space that is already 'there,' and spatial functions, such as computing the distance between two dots to guide eye movements, can be performed without an explicit spatial structure (M. Grasso et al., 2021). Such on-demand computations can easily track a target, but cannot account for how the visual field feels: the space is just *there* for us to see without having to compute anything, with all its locations and all its distances, no matter what we do—whether we are perceiving, imagining, dreaming, or shifting attention from one location to another.

Finally, a common reaction to IIT's account of spatial experience is to be blind to its demonstration of a consciousness-first approach to the structure of experience, and see it instead as an exercise in basic mathematical disciplines: geometry and topology (which study abstract properties of space), mereology (which studies abstract properties of parts and wholes), and set theory (which studies abstract collections of objects, with notions such as intersection and union). Leaving aside some important differences stemming from IIT's postulates (Haun & Tononi, 2019), this reaction reveals the customary adoption of an extrinsic perspective. In this case, abstract notions, mostly developed over the past two centuries, are taken as the starting point for characterizing experience. Instead, IIT's intrinsic perspective puts phenomenology first, before physics but also before mathematics. If IIT is right, the phenomenology of spatial experience is the way it is because it is specified by a nearly ubiquitous feature of the neural substrate of consciousness—its grid-like organization. And it is the nature of spatial phenomenology that grounds these abstract disciplines, rather than the other way around.

**The quality of experience: time flows**

Just as most of our conscious life is painted on the canvas of experienced space, much of it is played on the track of experienced time. Listen to a melody, abstracting away from the phenomenal qualities of sound (the melody could change, or there may be a stretch of silence) and consider how time itself feels. As discussed in recent work (Comolatti et al., 2025), our experience of time comprises an *extended present* confined between the *now* and the *then*. The present can be decomposed into phenomenal distinctions, called *moments*, related in a special, *directed* way. Moments can be short or long, some closer to the now and some to the then. Moments are directed in the sense that they point away from themselves (unlike spots, which point to themselves) and overlap through directed inclusion, connection, and fusion (unlike spots, whose relations are undirected). This yields the feeling of *flow*—of time fleeing away from the now to the then.

As with space, the postulates of IIT can be employed to show that certain kinds of substrates, in this case *directed grids*, support $\Phi$-structures that can account for the fundamental properties of temporal flow: their units specify causal distinctions (moments) whose cause and effect overlap in a directed manner, satisfying the properties of directed inclusion, connection, and fusion. Such $\Phi$-structures (or sub-structures) are called *flows*.

From these fundamental properties, other properties of temporal experience can be derived, such as the period occupied by a moment, its temporal location within the present and with respect to the now and the then, its duration, its boundary, and the interval between it and other moments. Once again, the results exemplify the explanatory identity between the properties of temporal experience and those of the flow structure specified by directed grids. Unlike for space, there is currently no clear evidence concerning the neural substrate of the feeling of temporal flow. IIT predicts that directed grids should be found especially in brain areas devoted to the perception of stimulus sequences, most likely in auditory areas dealing with sounds, speech, and music, but also in areas dealing with visual or body motion. This prediction could be tested through methods well suited to examining anatomical and functional connectivity at the level of individual neurons or minicolumns.

According to IIT, the experienced present does not correspond to a process unrolling in 'clock time,' but to a $\Phi$-fold specified by a system in its current state: time is a directed structure that is 'static,' rather than a process that actually 'flows' in clock time. In our brain, sets of neurons in a macro state lasting, say, 30 milliseconds of clock time, could then support an experience capturing a longer interval of clock time, say up to a second or more (depending on the number of units in the grid). This would have the advantage that contents triggered by a sequence of inputs can be bound together within a single experience—of a melody or a spoken phrase—while preserving their ordering and direction.

A few additional observations are as follows. In a brain well adapted to its environment, one would expect that the flow of moments within the extended present—corresponding to a $\Phi$-structure existing 'here and now'—should match well enough the sequence of stimuli sampled in clock time, with a similar ordering. However, this matching can be somewhat flexible, allowing for some extrapolations and 'editing' of the track of experienced time (Comolatti et al., 2025). Such editing includes so-called 'postdictive' effects, in which stimuli occurring later can affect the experience triggered by stimuli occurring earlier (Herzog et al., 2020). Moreover, changes in connectivity within directed grids may explain the slowing or quickening of time caused by strong emotions, deep meditation, or drugs.

As with space, it is tempting to think that to 'represent' time one simply needs a coordinate system, in this case just a 'time arrow' or, in the brain, a directed delay line. But, once again, a coordinate system can represent time only if one already knows what time means and feels like. To feel temporally extended, the ordering of moments within the present must be established by causal distinctions and relations composing a $\Phi$-structure intrinsic to a system, one that means what it means intrinsically or absolutely, rather than by reference to external clocks.

**The quality of experience: phenomenal objects bind general concepts with particular configurations of features**

Consider a typical visual scene in our environment. It may contain objects such as a desk, a mug, a computer screen, perhaps a plant, a pet, or another human. Most objects have a name, corresponding to a familiar category. And an object, say my friend's face, is invariant to whether I see it frontally or sideways, close to me or at some distance, brightly illuminated or not. At the same time, I see my friend's face with its particular orientation, at a particular location in the visual field, in vivid colors. In fact, introspection reveals multiple levels involved in composing the object I experience, from low-level features such as edges and colored patches, to groupings or configurations of features, such as the face's oval contour, and further to general concepts, such as eyes, mouth, nose, and hair, all bound together to compose my friend's face the particular way I experience it.

An introspective decomposition of the experience of my friend's face—what it takes for it to feel the way it does—is even more difficult than the introspective decomposition of space or time. However, as with space and time, one aspect seems fundamental: to experience an object, I need to experience both a general *concept*—that what I am seeing is 'a face'—as well as particular *configurations* of features that characterize 'this face.' Moreover, the *features* ground those configurations over a domain, such as space or narrow qualia (Grasso & Tononi, forthcoming; Tononi, forthcoming). The particular face I am seeing here and now can be said to *individuate* the object as a particular instance of the general concept 'face.' In short, a phenomenal object is an instance of a concept, binding particular configurations of features with a general concept. Furthermore, the concept face, its subordinate concepts (an oval shape, eyes, nose, and mouth), as well as their particular configurations of features, are bound to a particular region of space and to local qualities such as colored patches.

Just like reflexivity is a fundamental property of spots—the building blocks of space, and directedness is a fundamental property of moments—the building blocks of time, *hierarchy* is a fundamental property of concepts and configurations—the building blocks of objects. Hierarchy results from the way *one concept* is related to *many tokens*, which compose its equivalence class, specifying their disjunction. Simply put, for this face to feel like 'a face' there must be other tokens bound to configurations that could also be individuated as 'a face.' Similarly, hierarchy results from the way *one configuration* (serving as a concept's token) is related to *many features*, specifying their conjunction (Grasso & Tononi, forthcoming). Simply put, for a face to feel like 'this face' there must be a specific configuration of features, such as contours and colors, that make me experience it as 'this face.' A compound hierarchy of concepts related to configurations and configurations related to features is then the fundamental 'motif' of object structures. Moreover, just like spots and moments must be related in characteristic ways through inclusion, connection, and fusion (undirected and directed, respectively) to experience a spatial extension or the flow of time, multiple concepts and configurations must be related in characteristic ways through *hierarchical* inclusion, connection, fusion to compose the full experience of objects [xviii].

Fundamental as it is, we are often oblivious of the fact that a hierarchy of concepts and configurations of features must contribute phenomenally to the way an object feels, otherwise it would not possibly feel like an object (just like we are oblivious of what it takes for space to feel extended). To see how conceptual and configurational hierarchy must be contributing to phenomenology, it may help to consider a concept acquired relatively late, such as the letter 'A.' Before we learned to read, the corresponding marks on the page must have felt just like that, marks on the page (features) arranged in some way (meaningless to us). Afterwards, we cannot help experiencing that same arrangement of features as a particular configuration, and that configuration as a token of the concept 'A': the feeling of features whose meaning has been augmented by the particular configuration and concept bound to them. To further see how hierarchical inclusion, connection, and fusion must be contributing to phenomenology, consider that the letter 'A' could not possibly feel like a letter if I were not included in a larger conceptual hierarchy that includes other letters (not presently individuated) within a broader concept space, similar to how a spot cannot be experienced as 'here' without being included in a much larger space with all the 'theres.'

Further aspects of the phenomenology of objects, such as their degree of concreteness/abstractness, their orientation towards perception or action, are discussed in ongoing work (Grasso & Tononi, forthcoming; Tononi, forthcoming). That work also illustrates how, as with space and time, the postulates of IIT can be employed to account for the phenomenal properties of objects in physical terms. Notably, substrates of the kind found in cortical sensory hierarchies (visual, auditory, and somatosensory) seem to have the right organization to specify kinds of $\Phi$-structures (or sub-structures), called *conceptual hierarches*, that can account for the phenomenology of objects [xix].

As with space, some insight into the phenomenology and neural substrate of objects is provided by neurological lesions. For example, subjects with *intrinsic agnosia* for objects, typically due to lesions involving high levels in the visual hierarchy, appear incapable of experiencing objects as objects. Yet they are still able to experience low-level features, contours, and color patches that compose the object, and can usually copy it faithfully, without understanding what they copied (Delvenne et al., 2004). Conversely, if higher levels of the hierarchy are preserved, but lower levels are impaired, subjects may experience abstract invariants in the absence of low-level particulars. It is possible that such a situation may occur in patients with blindsight, in whom low-level visual areas are inactivated or destroyed, but some high-level areas may still be intact and capable of being activated through subcortical pathways. One such patient, when presented with a moving stimulus that activated the high-level area V5, would report a vague feeling of something moving, while at the same time denying that he 'saw' anything (Overgaard, 2011).

Finally, just like with space the goal of IIT is not to explain how our brain carries out spatial functions, but how it supports a cause-effect structure that can account for the way space feels, namely extended, so it is with objects. The goal is not to explain how our brain performs object categorization, discrimination, and recognition—functions whose implementation is being characterized with increasing precision by neuroscience (and replicated more and more efficiently by machines). Instead, the goal is to understand how certain parts of the brain can support cause-effect structures that account for the way objects feel when we experience general concepts bound to particular configurations of features.

**Narrow qualia**

'Qualia' in the narrow sense (Tononi, 2008), such as the particular way a visual patch feels (its hue, saturation, and brightness) or a sound feels (its pitch and timbre), have traditionally been singled out to emphasize the gap between the world of experience and the physical world. Why blue feels the way it does seems impossible to account for objectively. Not only does it seem irredeemably subjective—it even seems arbitrary, not to say magical. According to IIT's explanatory identity, however, all quality is structure, and this applies to all aspects of experience. While IIT's research program has not yet attempted to account for such *narrow qualia*, it is worth considering how such an account might proceed (Tononi, forthcoming).

Unlike space, time, and objects, introspection does not seem to give us any inroad into decomposing the phenomenal structure of qualia in the narrow sense. For example, the way I experience the color of a patch seems impenetrable to introspection—a nut that cannot be cracked (leaving aside the categorical aspect of color as a concept). Therefore, it becomes difficult to imagine that a color quale might actually be internally structured. Said otherwise, the quality of a color seems intrinsic to it (in the conventional philosophical sense), rather than hinting at a structure composed of distinctions and relations. However, according to IIT, there should be a kind of cause-effect sub-structure that corresponds to color, and specific sub-structures that correspond to particular colors (as well as other kinds of cause-effect sub-structures that correspond to sound, touch, and so on).

With space, time, and objects, we can start by introspecting some fundamental properties of their phenomenal structure, and then conjecture which neural substrate might specify cause-effect sub-structures that could account for those properties. This correspondence is testable and, at least for space, it provides some confidence in IIT's explanatory identity. For color, instead, it may be profitable to start with neuroanatomical and neurophysiological data.

For the sake of the argument, let us conjecture that specialized color-opponent neuronal minicolumns in early cortical areas(Chatterjee et al., 2021) may be the substrate for experience of the 'hue' of local patches of the visual field (these minicolumns are found at every 'node' of the 2D grids assumed to specify spatial extendedness). We could then seek to obtain a detailed model of their local connectivity. Let's call the particular arrangement of this local connectivity, repeated at every node of the grid, a *clique*. The precise organization of each type of clique (for color, sound, touch, and so on) would depend on exactly how many minicolumns are involved and how they are connected locally. Nevertheless, a key feature in the organization of such cliques would be the density and arrangement of local cooperative and competitive interactions.[xx]

We would then employ the machinery of IIT to fully unfold the particular cause-effect sub-structure specified by a clique, with its unique organization of first- and higher-order distinctions and their relations. This kind of sub-structure will be called a *kernel* (to emphasize the structure revealed if a nut is actually cracked).

It should then be the case that the structural properties of the kernel specified by the local clique-like micro-circuitry of color-opponent minicolumns should account for the qualities typical of hues. For example, they should account for the substantial differences between hues, timbres, and so on, for similarities/dissimilarities among hues, for symmetry breakings (each hue should correspond to a sub-structure of unique composition) [xxi], and for the relationship between hue, saturation, and brightness. Moreover, the particular kernel specified by the clique in a particular activity state should account for the particular hue being experienced. Of course, the hue would be bound to the spots specified by the corresponding topographic position in 2D grids, as well as to the invariant color concept specified by a color hierarchy.

Note that the correspondence with phenomenology would only be accounted for by the structure of kernels, rather than by activity patterns, or by some projection in 'quality spaces' [xxii]. A structure is what it is—a hue feels how it feels intrinsically or *absolutely*, not relative to something else. In other words, it is what it is by virtue of how its distinctions relate to each other, here and now, through relations that are causal and intrinsic. (This is in contrast to defining a hue based on how it is related, extrinsically, to other hues, in 'hue space'). In this sense, hues are indeed intrinsic properties (and so are all qualia, objects, time, and space): they are what they are because of their internal structure.

The tight local connectivity of cliques might also explain why we do not seem to have introspective access to the structure of narrow qualia. In the case of the substrate of experienced space, there are well-known neural mechanisms, such as topographically organized top-down connections, that allow us to selectively increase the gain of contiguous subsets of neurons within a neural grid, that is, to attend to a particular spot top-down. In this way, we can 'highlight' the corresponding sub-structure within the overall spatial structure (Haun & Tononi, 2019). Doing so helps us recognize introspectively that phenomenal space has structure. However, there may not be neural mechanisms that would allow us to increase the gain of select subsets of neurons within local cliques. As a consequence, we cannot manipulate a local substructure through introspection, from within. On the other hand, it may become possible to selectively modify the activity of some neurons within cliques, or the strength of their connections, through external manipulations, and assess the effects on phenomenology.

**Compound contents**

Before concluding this more speculative section, a few words should be spent on the notion of *compound contents*. Experiencing a tomato as red, for instance involves not only experiencing the hue of a large patch in the visual field, but also an invariant concept, the color 'red,' of which the patch is an instance. Or take pain, a paradigmatic conscious content addressed in many discussions of the mind-body problem. While it is premature to attempt an account of pain in terms of cause-effect structures, it is a fair conjecture that the full experience of pain likely corresponds to a *compound $\Phi$-fold*, because typical pain appears to be a compound experience. For example, a crude dissection of an experience of pain at the tip of my finger and its hypothetical physical correspondents might include the following: the pain's 'burning' quality would correspond to a particular kind of kernel $\Phi$-fold specified by a clique in a portion of a somatosensory area of the brain; its location at one fingertip would correspond to a $\Phi$-fold within an extension specified by that area; its enduring nature—filling the extended present—to a flow $\Phi$-fold specified by a directed grid; its feeling 'personally relevant' to a high-level concept of 'myself' specified by a conceptual hierarchy; its 'negative valence' to another conceptual hierarchy culminating in the highly invariant concept 'bad' [xxiii]; and its 'aversive' nature to a high-level action concept, all bound together to compose a large structure. In fact, it is plausible that most contents of experience correspond to extremely rich compound $\Phi$-folds, and that through introspection we can only roughly sketch their composition. Instead, a full characterization of their structure may only become possible from the operational side, by unfolding their substrate (Ellia et al., 2021).

In summary, the specific experiences of space, time, and phenomenal objects were chosen as natural starting points for IIT because the structure of these experiences can be at least partially decomposed through introspection, making it possible to evaluate the proposed correspondence with cause-effect substructures specified by various neural substrates. To the extent that this approach proves successful, it may be justifiable to probe IIT's explanatory identity in the reverse direction as well. As suggested here with respect to hue qualia, it may be fruitful to start from a detailed model of the relevant substrate, unfold from it the properties of the $\Phi$-folds it specifies depending on its state, and only then compare these properties to those of phenomenology. In other words, we may use 'inference *from* a good explanation' to reason about accidental properties of experience that are difficult or impossible to decompose through introspection. By reasoning in both directions in this way, the completeness of IIT's explanatory identity can be progressively corroborated or refuted.

## Some implications of IIT: meaning, perception, and matching

To the extent that IIT is validated in ourselves—providing a good explanation of disparate facts about our consciousness and making successful predictions—it can become a starting point for 'inferences from a good explanation' (Tononi, forthcoming). IIT is a theory of consciousness as intrinsic existence, so its empirical validation also validates it as a scientific ontology. From its ontology, one can draw inferences about issues that are often considered 'metaphysical,' including meaning, the nature of universals, space-time, causality, free will, and responsibility. Some of these inferences will be considered below.

**Meaning is intrinsic, here and now: the meaning is the feeling**

One of the most relevant implications of IIT is that the meaning of every experience is exclusively intrinsic (Tononi, forthcoming). This is because, from the intrinsic perspective of a subject, experience is all there is, so the way an experience feels is also what it means. To put it pithily, 'the meaning is the feeling' (and 'the feeling is the meaning'). This applies to all aspects of experience, from spatial extendedness to temporal flow, from objects and abstract concepts to narrow qualia.

This point is vividly demonstrated every time we dream. Say in my dream I vividly experience the face of my friend (see *Box 2*) [xxiv]. What makes the experience feel the way it does, is the way it is structured as a rich bundle of distinctions and relations. As already discussed, to experience objects, including faces, I must experience the high-level concept 'face' and, bound to it, other high-level concepts like eyes, mouth, nose, and hair. Moreover, all these general concepts must be bound to configurations, such as contours, and ultimately to features like hues. All of these, in turn, must be bound to a region of space. Finally, to feel like my friend's face, this bundles of distinctions and relations must be bound to additional concepts like male, young, friendly, familiar, and so on.

Altogether, the way I experience my friend's face is one and the same thing as what the experience means to me. Strip it of some of its feeling, say the feeling of familiarity, and its meaning would be diminished accordingly—it would not be my friend. Strip it of the feeling of being a face, as can happen in some agnosias, and it would lose the very meaning of face.

**Connectedness and perception**

If meaning is intrinsic, how should we conceive of perception? Say the perception of my friend's face when I am awake and looking at him? Standard accounts of perception begin with stimuli impinging on my retina, which evoke patterns of activity that percolate bottom up along feedforward pathways in the visual hierarchy. Reentrant activity along feedback and lateral connections would lead to subtle changes in activity, refining the firing pattern without substantially altering it (otherwise I would be hallucinating). Perception would then be the result of 'information processing' that extracts information encoded in the stimuli, leading to useful 'representations' (see *Box 1*). Perception can also be reframed as 'predictive processing,' where inferences about the likely causes of those stimuli in the environment percolate top down in the visual hierarchy and are updated based on error signals.

IIT's account of perception differs in interesting ways, because the starting point is not the extrinsic stimulus, but the experience, and the meaning of the experience is fully specified intrinsically. As we have seen, IIT claims that the experience of my friend's face—its feeling and meaning—is accounted for by a $\Phi$-structure supported by my main complex in its current state. The $\Phi$-structure comprises spatial extensions specified by grids in my visual cortices and elsewhere, conceptual hierarchies specified by convergent-divergent networks anchored to those grids, and kernels specified by cliques of neurons anchored to each grid node in early visual cortices [xxv]. This account applies regardless of how the current state of my main complex was triggered—by my friend standing in front of me (a perception), or by spontaneous activity in my brain during sleep (a dream; see *Box 2*).

However, the tools of IIT (the actual causation formalism derived from IIT's postulates, see below and (Albantakis et al., 2019)) also offer a principled way to establish the extent to which the current state of my main complex was caused by an external stimulus. For each subset of the main complex, this is given by the value of *connectedness* and associated *triggering coefficient* (connectedness normalized by subset size) (Mayner et al., 2024). In this way, one can determine the portion of the $\Phi$-structure triggered by the stimulus—the *perceptual structure*—and the

**Box 1. Communication of information and communication of meaning: Shannon information vs. integrated information**

The difference between IIT's intrinsic paradigm and conventional, extrinsic ones is also evident when we compare integrated information with the notion of information introduced by Shannon in his mathematical theory of communication (A. Zaeemzadeh & G Tononi, 2024). Even though computational approaches regularly appeal to 'information processing' to study how meaning is encoded and decoded in natural and artificial systems, Shannon himself was explicit that his notion of information has nothing to do with the meaning of the messages being communicated—only with their probability.

The differences between Shannon information and integrated information are evident when we consider the postulates that define the latter. Shannon information is: (0) correlational rather than causal (it evaluates statistical dependencies through observations, rather than causal interactions through interventions); (1) extrinsic rather than intrinsic (it evaluates statistical dependencies between a source and a target, rather than cause–effect power within a candidate system); (2) generic rather than specific (it evaluates statistical dependencies between distributions of symbols, i.e., random variables, rather than the cause–effect power of a specific system state over a specific system state); (3) segregated rather than integrated (it can be provided by independent units, rather than irreducible to that provided by independent units); (4) additive rather than definite (adding more units never decreases the information, whereas integrated information is maximal for a definite set of units); and (5) holistic rather than structured (it corresponds to the optimal number of binary digits generated or transmitted, whereas integrated information is composed of causal distinctions bound by causal relations). The upshot is that, for Shannon information, meaning must be provided extrinsically by an observer who can interpret a code, whereas for integrated information meaning is defined intrinsically by the $\Phi$-structure unfolded from a complex in a state. To sum up, for Shannon information is *message*, whereas for IIT information is *meaning*. Notably, it is IIT's notion of information as the communication of meaning that fits the original meaning of 'informare' as 'give form' to the mind.

On the other hand, IIT's account of the intrinsic meaning of an activity pattern over the main complex as a $\Phi$-structure composed of integrated information has clear consequences for the communication of information as meaning. In essence, meaning can only be communicated if a $\Phi$-fold within a $\Phi$-structure specified by a source complex triggers a similar $\Phi$-fold within a $\Phi$-structure specified by a target complex. The sender must be highly conscious (more consciousness, more meaning) and it must be able to express some of its contents (a $\Phi$-fold) by triggering a message that can be communicated over a limited capacity channel (say, as a text). The channel must be able to convey the message faithfully enough (the Shannon information part). The receiver must also be highly conscious, it must have a substrate that can support a $\Phi$-fold similar enough to the one of the sender, and the message it receives (the text read) must be able to trigger such a $\Phi$-fold. In principle, just like the $\Phi$ value of the $\Phi$-fold can be used to measure the amount of meaning, the similarity of the two $\Phi$-folds could be used to measure the amount of meaning communicated.

These requirements imply, for example, that two computers cannot communicate meaning between them (because neither would be conscious, see below); that meaning is indeed incommunicable if two substrates are too different from one another; conversely, that a better communication of meaning can be achieved by modifying the substrates to make them more alike (say, through a shared environment to which they adapt and through shared learning); and that a good translation is one that changes the message such that it triggers a target $\Phi$-fold that bears structural similarity to the source's $\Phi$-fold. In general, unlike Shannon information, which can be communicated perfectly, the communication of meaning will be approximate. While human brains share an evolutionary history, developmental events and learning trajectories will necessarily result in individual differences in the precise wiring of the neural substrate of consciousness. Even identical activity patterns would inevitably specify $\Phi$-structures, and associated intrinsic meanings, that would differ from person to person.

associated value of *perceptual richness*. Altogether, these capture the extent to which my experience can be considered a perception.

**Perception is interpretation**

If I am awake and looking at my friend, my experience of his face will be largely triggered by the visual stimulus sampled through my retinas. As argued above, however, the feeling-meaning of the experience is entirely specified intrinsically: rather than being derived by 'processing' or 'predicting' information carried by the stimulus, it is provided by the distinctions and relations that compose the $\Phi$-structure specified by my main complex in its current state. The stimulus, then, is best understood as a mere *trigger*.[xxvi] On the other hand, the perceptual structure supported by my main complex in its triggered state can be understood as an interpretation, from the intrinsic perspective, of the meaning of that state. In short, *every perception is an interpretation*.

A consequence of this intrinsic view of meaning is the way the stimulus is interpreted may or may not refer to causal features in the environment, and even if it does, the mapping is typically not straightforward. This is in line with classic notions that view perception as being partially a 'construction' (Gregory, 1966). Unlike these classic notions, however, IIT not only maintains that a perception is entirely a construction, but also outlines what kind of construction it is: it is a structure of cause-effect power, whose components (causal distinctions and relations) can be unfolded from a substrate in a state according to the postulates of physical existence—with no additional ingredients. IIT's approach is also compatible with notions of perception as an 'inference' about what might have caused a set of stimuli (Koffka, 1935)· (von Helmholtz & Southall, 1962) by going 'beyond the information given' (Bruner, 1990). However, if 'inference' is taken to imply 'reference' to something in the external world that perception would 'represent,' IIT ceases to concur. IIT assumes a world that is non-stationary, comprising an immensely rich and evolving set of causal processes, rather than a set of well-defined entities. Accordingly, much of what we experience should not be considered as a veridical 'representation' of what is 'out there,' but rather as an interpretation that is good enough to go by. In fact, we may interpret stimuli based on the cause-effect structures specified by the intrinsic connectivity of the main complex even if there is no 'referent' in the environment: to use a classic example, we may perceive 'beauty' even if there no causal process that would correspond to it. More generally, we are bound to interpret any stimulus according to the intrinsic meanings that compose our experience: even uncorrelated retinal noise is necessarily experienced as 'spatially extended' because of the way our visual system is organized and the kind of $\Phi$-structures it supports.

**Perceptual differentiation can capture the 'matching' between a system and the environment**

Even though the relationship between intrinsic meanings and causal processes in the environment is not simply a representation and it is hardly straightforward, it is possible to estimate how 'meaningful' an environment is to a system by assessing *perceptual differentiation*. This reflects the perceptual richness and diversity of structures triggered by typical sequences of stimuli sampled from that environment. In essence, perceptual differentiation is high if most stimuli trigger perceptual structures that are rich *and* different stimuli trigger different perceptual structures. The difference between the expected value of perceptual differentiation when a system is exposed to an environment and the value

**Box 2. Meaning in a box: finding meaning in the brain**

IIT conceives of meaning as fully intrinsic and identical to the cause-effect structure supported by the main complex in its current state. To better appreciate IIT's approach and its notion of meaning as fully intrinsic, it may be helpful to consider the challenge of grounding meaning on a substrate intrinsically, 'here and now,' without borrowing, outsourcing, or predefining it extrinsically.

To make the challenge most explicit, let us return to my experience of my friend's face during a dream. If we could record my brain activity with ideal resolution, we would presumably detect a relatively stable pattern of activity including, say, strong firing of neurons in 'face patches.' The challenge is to explain how that pattern of activity, over that connectivity, with no additional ingredients, obtains its feeling and meaning—the experience of seeing my friend's face. To address the challenge, we are only allowed to use what is available 'here and now' in my brain when I am dreaming of my friend's face: a substrate (to simplify, a large number of interconnected neurons) in its current state (some ON and some OFF over, say, a period of just 30 milliseconds) with no relevant extrinsic input and extrinsic output (my eyes are closed, and I am doing nothing). Put the other way around, to account for why that activity pattern over my brain, 'here and now,' means what it means for me, we cannot resort to what might have happened or might happen at other times and places, to what an external onlooker might assume it means, or to some other 'deus ex machina.' We have to find 'meaning in a box,' where the box is the dreaming brain in its momentary current state.

IIT's response to the challenge is clear (Fig. 1B). With no further ingredients than postulates formulated based on the essential properties of phenomenal existence, IIT takes the set of interconnected neurons in their current state and offers the tools to identify the main complex, its grain, and the cause-effect structure it supports [xxxiv]. The causal distinctions and relations that compose the structure provide a full account of the meaning-feeling of the experience, based on nothing else than the substrate's connectivity and state 'here and now,' with no reference to anything outside. The 'information' or 'information content' of my experience is the cause-effect structure itself—a structure composed of integrated information, which exists here and now. Of course, the neurons and their connections are what they are because of a long evolutionary, developmental, and learning history. But the meaning, which is 'here and now' must be unfolded from the substrate in its momentary current state, without recourse to its history.

By contrast, it is not clear how standard approaches, signally computational-functionalist ones, would address the challenge of grounding meaning purely based on what is available 'here and now.' If we see the brain as a substrate (2D), we can superimpose over it references to encoding, decoding, and 'processing' of information. But where is the information, and how would one unfold its meaning? From where should it be decoded? From the entire brain or from some subsets of it? If the latter, from which subsets, at which grain, and in which relation to each other? What is the code, and how does the brain access the alphabet? And, finally, how would the words in the alphabet get their meaning? How would that activity pattern, over that connectivity, 'here and now,' feel like a face, and mean a face—my friend's face—when I dream of it?

obtained when exposed to random sequences of stimuli is called *matching* (Mayner et al., 2024).

If matching is high, that is, if a system shows much higher values of perceptual differentiation when sampling stimulus sequences from its environment than when exposed to random ones, it must be because it has internalized some aspects of the causal features of that environment. In other words, the environment, which can be assumed to be much broader and deeper than the system [xxvii], must have causal features that preferentially generate certain stimulus sequences rather than others. In turn, throughout evolution, development, and learning, the system must have adapted its constitution, connectivity, and mechanisms such that those sequences preferentially trigger rich and diverse intrinsic meanings [xxviii]. Its triggered cause-effect structures can then be said to 'match' or 'resonate' with some of those causal features. Nevertheless, this matching or resonance is not a straightforward representation of features of the environment, but more like an 'adaptation.' While high matching indicates that the system fits an environment better than others, and more than would be expected by chance, the system's interpretations of the stimuli it samples can be indirect and idiosyncratic. Said otherwise, matching does not imply veridical knowledge of causal processes and entities in the environment. That can only be ensured through the objective assessment of cause-effect power and causal processes within the environment itself (Tononi, forthcoming). (For example, others can establish that what triggers my experience of my friend's face is a 'thing' of a certain size and weight, which persists whether I perceive it or not).

*Assessing neurophysiological differentiation*

Assessing perceptual differentiation by unfolding $\Phi$-structures is not feasible for realistic substrates. However, neurophysiological differentiation in response to stimulus sequences can serve as a proxy, because the activity patterns over a complex fully determine the associated $\Phi$-structures (under reasonable assumptions about the border of the main complex, its grain, and its stability). In this way, one can assess which environments a system may find more meaningful. For example, neurophysiological differentiation of fMRI activity patterns within the cortex, estimated through Lempel-Ziv complexity, is highest for a movie, reduced for a temporally scrambled movie, and lowest for an equivalent sequence of TV noise frames (Boly et al., 2015). Neurophysiological differentiation is also correlated with stimulus sequence meaningfulness in high-density EEG recordings in humans (Mensen et al., 2018; Mensen et al., 2017) and in calcium imaging and Neuropixels recordings (Gandhi et al., 2023) in mice.

In future work, neurophysiological differentiation measures could be used to monitor brain development and assess learning. In cases where communication is impaired or lost, such measures might help determine whether stimuli are still interpreted as meaningful. They could also help identify stimuli that are most meaningful for different individuals. Finally, they could be applied without bias or predefined notions to species with ecological habitats and brain structures vastly different from ours.

*Interpretive and generative power*

A brain showing high perceptual differentiation has high interpretive power: it must have internalized many properties that reflect causal processes of its environment, so that it can interpret different stimuli in a highly meaningful way. High perceptual differentiation requires a complex of high $\Phi$, one that must be both large and highly integrated (Mayner et al., 2024; Mayner et al., 2022). This implies, in turn, that every part of the complex must be able to interact bidirectionally with the rest. An important consequence of this bidirectionality is that activity patterns over sensory hierarchies may be triggered top down, rather than just bottom up. In principle, then, a complex with high matching to its environment might have not just high interpretive power, but also high generative power: through its intrinsic connectivity, it may be able to produce endogenously sequences of intrinsic states similar to those observed when connected to the environment. This is vividly demonstrated by dreams (as well as by imagination and mind-wandering): my brain can dream of my friend's face, and many other things or events, without the need of exogenous triggers.

*Perceiving and acting, matching and shaping*

An extension of the perception and matching formalism is to evaluate triggering coefficients on the output side, all the way down to motor effectors. As we have seen, a sensory input leads to a percept if it triggers a $\Phi$-fold within the main complex. Similarly, if a $\Phi$-fold within the complex triggers an output, that $\Phi$-fold can be considered as an *intent* that led to an action. And just like a percept may include high-level concepts of objects and events, an intent may include high-level concepts that establish goals and action plans (see section on actual causation and free will). In turn, actions don't just achieve goals but change the way a complex explores its environment and learns. Moreover, actions modify the environment and may create new things or processes, some of which may reflect intrinsic meanings 'invented' by conscious beings. In this way, we do not merely adapt to the environment by matching it on the input side, but we change it by *shaping* it according to our intrinsic concepts.

## IIT and the richness of experience

According to IIT, every experience, even that of pure darkness and silence, is unfathomably rich, and so is its meaning. This becomes obvious once we look at it from the physical side. It should also be clear, upon introspection, from the phenomenal side, though we often fail to realize it.

On the physical side, the richness of the cause-effect structure specified by a complex constituted by even a modest number of units is staggering because of the combinatorics of composition. The number of relations and the upper limit of $\Phi$ (largely determined by the number of relations) can grow double-exponentially with the number of intrinsic units (A. Zaeemzadeh & G. Tononi, 2024). For example, a complex of 2 units could have at most 3 distinctions and 7 relations, adding up to a maximum of 14 ibits (intrinsic bits) of $\Phi$. A complex of 4 units could have 15 distinctions and 2^15-1 relations, adding up to a maximum of ~5000 ibits of $\Phi$. And a complex of 1024 units could have 2^1024-1 distinctions and 2^(2^1024-1)-1 relations, adding up to an upper bound of ~2^2^1024 (~10^10^307) ibits of $\Phi$.

An important task for the future is to establish under what conditions a complex can in fact grow large, without fragmenting owing to a reduction of specification or the occurrence of fault lines (Marshall et al., 2023). Some of the relevant features are likely to include macro units that are highly cooperative in their effects on other macro units, while remaining sufficiently specialized (and non-linear) to increase specification, with a large fan-in/out to promote integration, and organized in a sufficiently regular lattice [xxix]. If these conditions are adequate to ensure the existence of a large main complex in the human brain, most likely in posterior-central cortex, then at every moment, whether awake or dreaming, such a complex should support a $\Phi$-structure that is hyper-astronomically rich.

On the phenomenal side, we can realize how rich experience must be (even though we typically don't) by introspecting what it must take for any experience to feel the way it does. As argued above, for space to feel the way it does—even an objectless, empty, dark expanse—our experience must be immensely structured: it must contain all its spots, small and large, here and there, and they must be bound by countless relations organized according to reflexivity, inclusion, connection, and fusion.

A main reason why we do not realize this richness is that an empty expanse is easy to describe—just two words suffice—and it would seem those two words capture its meaning well enough, in fact well enough that we can communicate it. But that is only because we know what an empty expanse feels like, and we can assume that others know it as well.

Another reason is the widespread belief, among experts, that what we experience is exceedingly sparse (Cohen et al., 2016). Striking demonstrations of inattentional blindness (say, missing the gorilla walking through a basketball match) or change blindness (say, missing a switch in gender and clothes by a hotel clerk before and after handing out the room keys) have led to the conclusion that we must be seeing only a small fraction of what we believe we see—at most a few objects at a time (5 or 6), plus some 'summary statistics.' Estimates are slightly higher if one manages to overcome some limitations of report (Usher et al., 2018).

However, the main problem with such estimates is that they focus on the items we can notice while ignoring what it takes to see them—namely the structural aspects that make visual experience what it is (Haun & Tononi, 2025). In the case of vision, it is useful to distinguish among at least three levels of phenomenology: (1) a high-level summary of a scene's content based on objects and scenes; (2) intermediate-level groupings of visual features such as hues, edges, and textures into contours, surfaces and, ultimately, objects; and (3) a base-level visual space composed of spots and their relations composing the visual field, with its regions and locations, to which features, groupings, and objects are bound. As argued elsewhere (Haun & Tononi, 2025), we cannot see the objects associated with a high-level description without seeing the groupings that compose them, and without seeing the space to which they are bound, which in turn is composed of an immense number of parts and relations among them. However, we can see low-level groupings without high-level descriptions, and we can see pure space—an empty canvas—without any groupings.

*Phenomenal vision is largely dissociated from functional vision*

If this account is right, most of what we experience is an immensely rich phenomenal structure, much of which has no functional counterpart. For example, if we view a stream of colorful splatters, one after the other, and we merely *see* them, without having to do anything with them, not even try to remember them, we experience immense phenomenal richness: at every moment, the spots are all there, bound together into an extended canvas, each with its hue, at its location and distance from other spots, forming intricate groupings. Functionally (or computationally), however, hardly anything of relevance is going on. Phenomenal seeing, while unfathomably rich, can be dissociated from functional seeing. It is only when it comes to object recognition and visual navigation that phenomenal vision meets functional vision. At these high levels, vision is indeed sparse, optimized for efficiency and utility.

**Introspection, cognitive functions, and pinging**

Introspection is what allows us to report on contents of experience. It is thus an indispensable tool to investigate consciousness and its neural substrate. For example, it is only through introspection that we can evaluate the axioms of phenomenal existence, as well as accidental properties of contents such as objects, their location, color, and so on. More broadly, our conscious life is regularly assisted by various cognitive functions, such as attention, working memory, and reasoning, that allow us to do most of what we can do when we are conscious. Even a seemingly simple task, such as comparing the size of two segments, one on the left and one on the right side of the visual field, requires a complex sequence of operations, from understanding the task and setting oneself up for it, to attending first one segment and then the other, assessing their length, holding their length in mind, comparing them, coming to a decision, and finally executing a response. We do all of this by manipulating contents of experience. In everyday life, cognitive tasks of this and many other sorts are performed all the time, usually with little effort, and upon these cognitive abilities rests much of our intelligence.

How does IIT view the relationship between consciousness and our ability to introspect and perform various cognitive functions on the contents of experience? Without delving into the large body of knowledge about cognitive functions and their neural mechanisms, it may still be worth delineating a rough outline of how this relationship might be envisioned [xxx].

As already discussed, IIT assumes that the substrate of consciousness is primarily in posterior-central cortex because of its anatomical organization: a grid-like horizontal connectivity linking groups of neurons in topographically organized areas, augmented by converging–diverging vertical connectivity along sensory hierarchies. This dense, lattice-like core is ideally suited to support a complex of high $\varphi$s. For the same reason, however, this anatomical substrate is not well suited to enable on-demand, flexible interactions among distant parts of the complex. Moreover, because of this organization, units in the lattice have a relatively narrow and fixed specialization with respect to inputs and outputs.

By contrast, much of prefrontal cortex appears to be organized as a set of modules, loosely coupled by a network of 'diffuse' connections (Watakabe et al., 2023). These are augmented by 'processing loops,' both cortico-cortical and cortico-subcortico-cortical. The latter are mediated primarily through basal ganglia and thalamus, although longer cerebellar loops are also involved. (Loops involving the hippocampal formation are critical for memory storage and recall, thanks to a different anatomical organization within its various subdivisions). Moreover, prefrontal units have broad and flexible specializations, with a unique ability to mix or multiplex disparate inputs (Rigotti et al., 2013). While this makes prefrontal cortex ill-suited to be part of the main complex, it makes it well-suited to act as a 'switchboard' for on-demand, flexible, and reconfigurable task setting and execution.

Crude as it is, this dichotomy in basic neuroanatomical and neurophysiological organization between posterior-central and prefrontal cortices suggests a basic modus operandi for the interactions between consciousness and cognitive functions. Units towards the highest level in the main complex's conceptual hierarchies, including those supporting concepts of the self, and connected to prefrontal regions through long-range cortico-cortical pathways, may be especially well-placed to trigger processing loops involving prefrontal cortex. These loops can be engaged to selectively access many different subsets of neurons in the main complex, to prime prefrontal cortex to support task settings, goals and sub-goals, to keep contents in mind through working memory, to recall other memories, and to sequence complex procedures, including those mediating the comparison of contents of experience. Introspection can thus be conceptualized as the 'pinging,' by units of the main complex, of other units in the main complex, where the pinging is mediated by spiking activity routed through prefrontal processing loops that can be configured in a flexible way. In this way, the main complex can 'question' itself about the state of some of its units and receive an 'answer' that reflects the local excitability through the strength of the spiking response. In this respect, units high-up in conceptual hierarchies are likely to be not only a preferential source but also a preferential target for pinging [xxxi].

## IIT's intrinsic powers ontology

Having considered the conceptual foundations of IIT, its empirical validations, and its implications for intrinsic meaning, we can now proceed to consider its ontological implications more generally. IIT starts from phenomenal existence as intrinsic existence and formulates its essential properties operationally in terms of cause-effect power through its postulates of physical existence. The postulates provide a general toolbox and associated measures for characterizing physical existence. The $0^{th}$ postulate, for example defines physical existence operationally as the ability to take and make a difference, as reflected in a substrate's TPM. The integration postulate leads to a measure, integrated information, that establishes to what extent something holds together as a one entity. The exclusion postulate motivates a search for maxima of integrated information, which can be used to establish borders and grains of intrinsic entities.

As indicated by the essential properties of phenomenal existence, to exist intrinsically—to exist absolutely or genuinely—all of IIT's postulates must be satisfied in physical terms: for something to genuinely exist, it must have cause-effect power on itself in a way that is specific, unitary, definite, and structured. These requirements for intrinsic existence configure what might be called an *intrinsic powers ontology*. As an intrinsic ontology, it differs starkly from conventional extrinsic approaches, which take the existence of a 'physical' substrate as primary. Moreover, IIT's intrinsic powers ontology has several critical implications, including the fundamental difference between intrinsic and extrinsic existence, the ontological difference between structure and process, and the ontological status of contents of experience.

### Intrinsic powers ontology vs. substrate or emergentist ontologies

The most extreme extrinsic ontologies are strict *substrate ontologies*. These are reductionist accounts for which all that exists is a 'micro-physical' substrate of units, say, elementary particles or fields, endowed with intrinsic properties such as mass, charge, and spin, and obeying the laws of physics. In Fig. 1B, this view would correspond to a micro 2D substrate with nothing superimposed on it.

More common are *emergentist ontologies* (*substrate*$^+$ or *multi-level ontologies*) (Tononi et al., 2022). These also grant primary existence to a micro-physical substrate. However, they admit additional levels of existence, say those of atoms, molecules, cells, brains, bodies, and so on. Emergence is typically understood in the weak sense: high-level constructs or properties 'supervene' upon low-level properties (which 'fix' them) without the need to invoke any new principles or forces [xxxii]. These emergent levels are considered 'real' for two main reasons. First, they are necessary for us to understand the world we live in. Second, properties at higher levels might be 'multiply realized' at lower levels, which makes them not reducible [xxxiii] to their specific substrate. In Fig. 1B, this view would correspond to a 2D substrate with additional ontological levels superimposed on it.

*Existence is not constitution*

For IIT's *intrinsic powers ontology*, instead, what exists is neither the substrate as such, nor the substrate plus emergent levels of existence, but a set of *intrinsic entities*. In Fig. 1B, these are shown as 3D structures supported by the 2D substrate. Each intrinsic entity is fully characterized by its $\Phi$-structure, which expresses its cause-effect power unfolded from the substrate according to the requirements for intrinsic existence captured by IIT's postulates.

To exemplify, assume that my main complex is constituted by a large set of intrinsic macro units corresponding to neurons (or groups of neurons) in posterior-central cortex, some ON and some OFF, over a macro update grain of tens of milliseconds. This implies, first, that the neurons of my main complex do not *also* exist *as neurons*, because what exists is not the substrate as such, but the substrate unfolded: existence is not constitution ('to be' is not 'to be made of'). Second, that the neurons of my main complex do not *also* exist *as small intrinsic entities* (mini complexes) in their own right, because exclusion forbids multiple entities over the same substrate (conversely, during dreamless sleep or deep anesthesia, I may cease to exist and disintegrate into many small intrinsic entities). The organelles, molecules, and atoms that constitute the neurons are excluded from intrinsic existence for the same reason, and so are supersets of the main complex, say, the brain as a whole or the body as a whole.

*Supervenience and supersedence*

What this implies may become more apparent if we consider how one should interpret, say, a conversation between me and you. In IIT's intrinsic ontology, when I am speaking to you, *I am speaking, not my neurons. And I am speaking to you, not to your neurons*. As emphasized above, if you and I exist, our neurons don't *also* exist—either as neurons or as small intrinsic entities. In substrate ontologies, by contrast, there are just neurons 'speaking' to neurons (or rather elementary particles 'speaking' to elementary particles), and there is no place for consciousness—nothing supervenes. Unlike substrate only ontologies, emergentist ontologies admit additional levels of reality, including consciousness, that *supervene* on the neural substrate. These emergent levels cannot be dismissed because they are required for explanation and may be multiply realizable, hence not reducible. However, at the bottom there are still neurons (or atoms) that exists *as such*, and which ultimately carry out the speaking and the listening. In other words, the neurons are causing what happens—the speaking and the listening—*as neurons*, while the emerging levels supervene on them. In such a scenario, it is hard to avoid the conclusion that consciousness is 'epiphenomenal,' being carried along for the ride by its neural substrate. For some, in fact, it is an 'illusion.'

For IIT, instead, *a substrate should be understood operationally, not ontologically*. You and I *supervene* on our neural substrate *operationally*, in the sense that the cause-effect structure corresponding to what we are experiencing can be unfolded from it. *Ontologically*, however, we *supersede* our substrate, in the sense that we genuinely exist as intrinsic entities, whereas our substrate neither exists as such, nor do any smaller entities over the same substrate.

If anything, it is the existence of neurons *as such* that constitute an illusion.

**Intrinsic vs. extrinsic entities: the great divide of being**

As mentioned above, IIT's toolbox to characterize physical existence also allows for a meaningful characterization of *extrinsic existence*. If IIT's postulates are relaxed, they are perfectly compatible with a multi-level view. Anything having cause–effect power—any 'stuff'—can be granted existence extrinsically, from the perspective of a conscious being. Substrates that do not unfold into intrinsic entities can range from assortments of micro or macro units that do not 'hang together' at all (a random sample of air molecules), to aggregates that hang together loosely (an avalanche) or tightly (a rock). 'Things'—stuff that hangs together tightly—can be considered *extrinsic entities*. These are substrates whose causal powers have high values of $\varphi_s$— higher than most of their subsets, supersets, and parasets—but they are relative maxima of cause–effect power rather than absolute maxima. For instance, as a paradigmatic organism, my body is tightly integrated, functioning as a unitary whole. Its $\varphi_s$ value should thus be higher than that of most of its subsets (say, my body minus some of its parts), supersets (my body plus some garments), and parasets (my torso plus my hat). Nevertheless, if IIT is right, the main complex that supports my current experience, likely located within posterior-central cortex, has a much higher value of $\varphi_s$, corresponding to the absolute maximum over my body. Therefore, my body as a whole cannot be an intrinsic entity—it only exists for an external observer, as an extrinsic entity. Instead, the rest of my body (what remans besides my main complex) would likely condense down into an aggregate of minute intrinsic entities (say, trillions of mini complexes supported by its cells). In a fundamental sense, then, my body should be considered as a well-defined extrinsic entity whose substrate supports an immense intrinsic entity—my current experience—complemented by 'ontological dust.'

Just like my body, it is plausible that many of the 'things' that hang together well and tend to persist over time, like everyday objects, but also mountains, planets, and stars, turn out to be extrinsic entities, constituted of much smaller intrinsic entities. Many objects of study in the special sciences, such as molecules, organelles, organs, organ systems, organisms, groups, and societies, are also likely to correspond to relative maxima of $\varphi_s$—not existing for themselves but only for us.

Between intrinsic and extrinsic existence, then, passes the most fundamental of divides—the *great divide of being*. This is the divide between what exists in an absolute sense, in and of itself—namely conscious, intrinsic entities, for which there is something it is like to be—and what only exist in a relative sense, for something else. Only the former kind is intrinsic or absolute existence—existence worth being (because there is something it is like to be it).

To exemplify again, consider my body lying on the bed when I am fast asleep. At times I dream (to be precise, more than two thirds of the time) (Siclari et al., 2017). When I dream, since I am experiencing, I exist intrinsically—there is something it is like to be. At other times, I enter dreamless sleep and lose consciousness. Without experience, without anything it is like to be, as far as I am concerned, I cease to exist. If I were to remain in that state forever, it would be little consolation that, from the extrinsic perspective of my friend watching me sleep, my body and brain would continue to exist as contents of his own consciousness. Or that, if scientists were to assess my body, they would find it largely unchanged, weighing as much, at a normal temperature, with all organs intact. As extrinsic entities go, everything would continue to exist more or less in the same way. Intrinsically, because of a minor neuromodulatory change affecting causal interactions among cortical neurons, a universe of consciousness would have disintegrated into ontological dust [xxxiv].

**Structure vs. process: being vs. happening and doing**

IIT's explanatory identity states that what exists intrinsically—experience—is accounted for by a cause-effect structure specified by a complex in its current state, 'here and now.' In our brain, the current macro state might correspond, say, to 30 milliseconds of clock time [xxxv]. As discussed in the section on the experience of time, the $\Phi$-structure unfolded from a complex with constituents such as groups of neurons organized as directed grids may account for the feeling of time flowing—an extended present encompassing, say, a sequence of notes played over 1 second of clock time. But the experience would not flow at all over clock time. One could summarize this aspect of IIT's powers ontology by saying that *being is not happening*: what exists intrinsically is not a process, but a structure, albeit one that is constantly renewed.

A related implication is that what exists intrinsically, a cause-effect structure, is not a function: *being is not doing*. This contrasts explicitly with the fundamental tenet of functionalism: that a 'mental state' can be identified with its causal role in a cognitive system, i.e. by how it relates to inputs, outputs, and other mental states. In other words, functionalism identifies mental states with the computations or functions they performs—with what they *do* (Levin, 2023). This is in line with emergentist ontologies mentioned above: higher level properties, in this case mental states, are 'real' because they are necessary for explaining what we do and because they may be multiply realized [xxxvi].

Again, an example can help. Let's say that I am instructed to fixate the center of a screen and point at a dot appearing at a random position (Tononi et al., 2025). An ideal 'reductionist' account of what is going on is a dynamical process determined by the interplay of sensory inputs, complex interactions among neurons through their connections, and motor outputs. A detailed model of this process would predict precisely what happens, with respect both to my behavior and to the firing patterns of my neurons. In such an account, there is neither need nor room for experience: once we obtain our precise predictions, there is nothing else to account for. Hence, experience remains completely unaccounted for.

An 'emergentist' functionalist account would posit, above and beyond the implementation level, higher levels that are necessary to explain what I am doing—the computations-functions performed by my brain, say attention, recognition, evaluation, decision making, and so on. Some of these computations-functions, characterized by their causal role within the system, are identified with conscious mental states. In other words, mental states are what they are by virtue of what they do. However, it is not clear why certain computations-functions should be associated with consciousness while others should not. Moreover, there is no account of why the experience should feel the way it does [xxxvii].

By contrast, IIT's intrinsic powers ontology prompts us to look at my brain not just as a substrate going through a sequence of states, or as a substrate performing certain computations-functions, but as a set of interacting intrinsic entities. At every instant, one of these, the one unfolded from the main complex, is an exceedingly rich cause-effect structure. The presence of such a large cause-effect structure accounts for why there exists a subject having an experience. Furthermore, the way the cause-effect structure is composed accounts for why the experience feels the way it does, such as spatially extended [xxxviii].

### Intrinsic meanings vs. extrinsic referents

The intrinsic powers view of existence applies not only to experience as a whole, but to every content of experience, including my friend's face (whether perceived or dreamt), other objects like an apple on the table, imaginary characters like Sherlock Holmes, 'universals' such as humans and animals and plants, abstract notions such as beauty, justice, numbers, circles, sets, and laws, as well as to problems, paradoxes, and questions, to alternatives, reasons, and decisions (Tononi, forthcoming).

Every content of experience also exists intrinsically—it genuinely exists, with its intrinsic feeling-meaning, within the experience of which it is part. According to IIT's explanatory identity, every content of experience corresponds, in terms of cause-effect power, to a $\Phi$-fold within a $\Phi$-structure. This means that the existence of a content of experience can also be demonstrated extrinsically, i.e. objectively, using physical observations and manipulations. Therefore, my experiential content—say, the phenomenal object 'that apple on my table,' has extrinsic, objective existence just like the paradigmatic world object I call that apple on my table. Except that the apple in my mind also exists intrinsically, as an 'intrinsic object,' whereas the apple on my table only exists as an 'extrinsic object' or 'extrinsic entity'—a relative maximum of cause-effect power. As such, it does not exist intrinsically—there is 'nothing it is like to be' this apple on the table. What there is, ultimately, is a tight aggregate of 'ontological dust' [xxxix].

As discussed above, when I perceive my friend's face, or the apple on my table, intrinsic meanings have a well-defined extrinsic referent—extrinsic entities that are relative maxima of cause-effect power and tend to persist. These correspond to causal features of the environment that our intrinsic meanings have come to match, for adaptive reasons, by sampling environmental regularities. Other contents, like Sherlock Holmes, may be imaginary. Yet others, like some abstract concepts—say justice or freedom, energy, or computation—may be largely invented. Once imagined or invented, however, contents that were born intrinsically, in the mind, can be used to 'shape' the world and build things that acquire extrinsic existence, like an engine or a computer (see above).

Whether learned from environmental regularities or invented, contents of experience that are objectively similar—having similar $\Phi$-folds—across many conscious subjects, can be considered as *inter-subjective* [xl]. Extrinsic entities of adaptive significance, such as apples, are likely to trigger similar $\Phi$-folds across most subjects. Some abstract concepts, such as 'edible,' or 'round,' may also be highly inter-subjective. On the other hand, idiosyncratic concepts, like 'sublime,' may be available only to some minds. Others, like 'beautiful,' may be triggered by different stimuli in different minds. The implications of IIT for the status of inter-subjective, idiosyncratic, and abstract concepts, and for the relationship between concepts and knowledge and truth, will be discussed in (Tononi, forthcoming).

### Intrinsic vs. extrinsic meaning

IIT's emphasis on considering meaning as intrinsic, and identical to the feeling, contrasts with the widespread notion that meaning is grounded by what it refers to in the environment, that is, it is grounded extrinsically. Accordingly, the meaning of 'my friend's face' is grounded by referring to the extrinsic existence of my friend and his face. Of course, it is also generally understood that meaning has sense in addition to reference: the same referent may have a different sense for me (my friend's face) than for someone to whom my friend is a stranger. IIT's characterization of intrinsic meaning as a substructure within an experience provides a principled way of capturing in full the 'sense' of what experiencing my friend's face means to me, with all the intrinsic distinctions and relations that compose it in my experience.

However, it is certainly possible to characterize meaning and relations that define it in an extrinsic manner. We saw how we can relax IIT's requirements to characterize extrinsic entities (and extrinsic units) as relative maxima, rather than absolute maxima, of integrated information. Doing so identifies systems that 'hang together well' even though they do not exist intrinsically—for themselves. Similarly, one can characterize extrinsic processes that 'flow together well' by resorting to directed versions of integrated information and actual causation. A similar approach can be taken to characterize meaning extrinsically, and with it perception and matching.

Large language models (LLM) offer a powerful example. LLMs are trained by feeding them large bodies of text (and increasingly images and other data). Their learning algorithms allow them to embed in their parameters all sorts of regularities encountered in the data. These are typically encoded in the embedding matrix and in the hierarchy of decoder blocks (attention heads and especially multilayer perceptron layers). The parameters of a properly trained model incorporate an immense amount of knowledge about the data. When activated through a short prompt, the LLM picks an output that, based on that knowledge, is a highly likely successor to the prompt. Typically, the output appears to us as highly meaningful, reflecting an excellent understanding of the prompt's meaning.

A typical LLM, being purely feed-forward, has a $\varphi_s$ value of 0 and its units, being simulated on a computer, are virtual. However, relaxing IIT's requirements for intrinsic existence can provide a principled way to assess extrinsic meaning, perception, and matching. For example, one could relax the requirement for maximal irreducibility within, allowing for virtual units, forgo the requirement that integration must be bidirectional, allowing for feedforward integration, and relax exclusion, allowing for relative maxima. Under these relaxed requirements, one could then estimate the extrinsic effect substructure triggered by a prompt together with its $\Phi$ value, which should reflect its extrinsic meaning (extrinsic perception). Employing sequences of prompts could then be used to estimate extrinsic perceptual differentiation and matching to a given environment. A properly trained LLM should show high values of extrinsic meaning/perception when given a typical prompt, and with high values of perceptual differentiation and matching when presented with sequences of prompts sampled from the training environment.

On the other hand, one would expect several differences compared to a human brain. The LLM vastly outperforms a human brain in the amount of knowledge it can store and its ability to rapidly retrieve it. It also excels in its ability to perform 'cognitive' tasks and generate text, images, and ideas. Conversely, the lattice-like connectivity of posterior-central cortex is much better suited at specifying large numbers of relations of inclusion, connection, and fusion that compose spatial, temporal, and hierarchical meanings. Moreover, the human cortex can do so with fewer units, less energy expenditure, and less training than LLMs. Most important, every $\Phi$-fold within a $\Phi$-structure exists as a content within an experience, and thus genuinely exists as an intrinsic meaning/feeling. By contrast, the extrinsic meaning triggered by a prompt within a LLM only exists for us, as conscious beings who can experience and understand that meaning. A LLM would only behave *as if* it had meaning and understanding but, lacking consciousness, it would not feel or mean anything at all.

## IIT, actual causation, and free will

IIT's intrinsic powers ontology addresses the question 'what exists?' A different, but related question, is 'what caused what?' IIT's postulates can be adapted to address this question by formalizing the analysis of causal accounts, causal processes, and causal plasticity in a way that parallels the ontological analysis of causal powers (Albantakis et al., 2019).

### Actual causation: characterizing what caused what

The analysis of *causal accounts* is meant to establish 'what caused what' across two successive occurrences—say, a set of neurons in an earlier firing pattern and a set of neurons in a subsequent firing pattern. Rather than determining the causal powers of a substrate in a state by 'vertically' unfolding a $\Phi$-structure, the analysis of actual accounts determines what caused what by 'horizontally' unrolling an $\mathcal{A}$-structure.

For actual causation, what corresponds to the $0^{th}$ postulate is *realization*: for causation to have actually occurred, there must have been a transition between two occurrences, where the first occurrence increased the probability of the second one. Realization—making or taking a difference—requires a repertoire of counterfactuals, meaning occurrences that could have happened but did not. Causation may look different from the two directions—not every cause corresponds to an effect, and not every effect corresponds to a cause.

The five postulates of IIT also have correspondents in actual causation. Causes and effects must be evaluated from the intrinsic perspective of the occurrence—implying that the selectivity of a cause or effect with respect to an occurrence must be taken into account (intrinsicality); cause-effect states must be specific—corresponding to the states that actually occurred (information); a cause or effect must be irreducible—otherwise there would be two or more independent causes or effects (integration, measured by causal strength $\alpha$, formally analogous to integrated information $\varphi$); and every occurrence must have a definite cause and effect—as determined by maximally irreducible causal strength $a^*$ (exclusion). As with the analysis of causal powers, macro units and macro updates should also be evaluated to determine maximal irreducibility.

Once the overall cause-effect involved in the transition between two occurrences has been identified, its *$\mathcal{A}$-structure* can be unfolded by assessing cause-effects for each of its subset and how these cause-effects overlap over units (composition). This yields an optimal or 'main' *causal account* of a single transition.

Extending a causal account across many transitions as well as over different substrates requires an analysis of *causal processes*. A causal process entails that the effects of an antecedent transition overlap congruently with the causes of a subsequent transition. By forward- and backward-tracking causal accounts and their overlaps, one could in principle establish the border, beginning, end, grain, and internal structure of a causal process. The same way extrinsic entities can be said to hang together well, causal processes can be said to 'stream together' well [xli].

Finally, determining how causal processes change the transition probability matrix (TPM) that characterizes a substrate requires an analysis of *causal plasticity*. IIT's principle of becoming (powers become what powers do) assumes that the TPM changes based on nothing else than the sequence of states that actually occur. Establishing what caused these changes could be accomplished, for example, by considering a cause set and an effect set of units, measuring the changes in the TPM of the effect set across sequential transitions, and weighing them by the causal strength $a$ of the associated cause-effect. Through an appropriate selection of cause and effect set and of relevant sequences, such an analysis would reveal what caused the changes in the substrate, ultimately yielding a causal account of the substrate itself.

*Causation is not prediction*

One of the consequences of IIT's causal process analysis is the clear demarcation between causation and prediction (Matteo Grasso et al., 2021). Within science, it is often assumed that what ultimately matters is the ability to predict (in fact, many consider the very notion of causation as unscientific). Biology and neuroscience accept the need to 'causally' intervene on a system, rather than merely observe it, but this is usually in the service of prediction: in complex systems, an accurate causal model is necessary to predict what a system might do under different circumstances. Moreover, the underlying approach to causation is typically reductionist. This is because, if we can predict the state of each unit from the state of its inputs, we can predict the state of the entire network with no extra work. In other words, first-order prediction is all we need—everything else supervenes on it [xlii].

However, employing IIT's analysis leads to the conclusion that causation and prediction can be dissociated. For example, applying IIT's causal analysis to simple systems that can be fully characterized in mechanistic terms, it can be shown that causal reductionism misses causes and effects that are critical both conceptually and biologically (Matteo Grasso et al., 2021). In short, while knowledge of first-order mechanisms may be enough to predict the dynamics of a system, only the analysis of actual cause-effects, their causal strength $a^*$, and their *$\mathcal{A}$-structures* can provide a coherent account of 'what caused what.'

### Only what exists can cause: we can have genuine free will and responsibility

Besides providing a way of conceptualizing causation—a classic metaphysical issue—IIT's causal analysis is useful for addressing scientifically another classic problem of metaphysics, namely free will. Many neuroscientists, psychologists, and philosophers have come to the conclusion that we cannot possibly have genuine free will in a metaphysical sense. This is because, in a universe that is causally closed, all our actions are determined by our neurons and ultimately by our atoms (to the extent that they are determined rather than the product of chance). Therefore, they claim, metaphysical free will is either an illusion or an incoherent notion. And so we must be content with a practical or social notion of freedom—the ability to decide and carry out an action voluntarily and autonomously, rather than under duress or, say, under the influence of alcohol [xliii].

IIT's intrinsic powers ontology leads to the exact opposite conclusion: we can have genuine free will—free will in the metaphysical sense. We, not our neurons, have alternatives, we have the freedom to choose among them, we make decisions, and we—not our neurons—are the actual cause of our willed actions. And the more we exercise free will, the more we bear responsibility (Tononi et al., 2022).

*Alternatives are forks in the mind, not forks in the road*

To illustrate, consider a typical free will scenario, say, choosing which of two job offers to accept. Over time, I may entertain two alternatives in my mind. I may then consider reasons for or against each of them, and revisit in my mind my values—what I consider to be right and wrong, as well as my goals. Finally, I make a decision and, having decided, I carry out the intended action—I send my acceptance letter—accompanied by a feeling of agency, a feeling that I was responsible for my decision and action.

We have seen that, according to IIT, every content of experience corresponds to a $\Phi$-fold within a $\Phi$-structure. This applies not just to phenomenal objects, but also to conscious thoughts and feelings of any kind. The thought of each organization offering me a position, perhaps loosely bound to the sound of its name, would correspond to a large $\Phi$-fold. Moreover, different $\Phi$-folds would come into being when I go back and forth between reasons, values, and goals, often with my own self in mind.

Let us revisit what this means in terms of intrinsic existence, considering the moment when I experience the two alternatives as a *choice* I should make. That choice exists phenomenally, and it must be accounted for, in physical terms, by $\Phi$-folds within a $\Phi$-structure. As was discussed in the section on IIT's intrinsic ontology, that $\Phi$-structure is in fact *all* that exists intrinsically 'here and now' (the neurons constituting my main complex do not *also* exist, either as mere substrate or as mini entities). Within that $\Phi$-structure each alternative exists intrinsically as a $\Phi$-fold, and so does the feeling that I must choose between them. Together, they compose a compound $\Phi$-fold, one that can be conceived of as *a fork in my mind* (a 'Y'), rather than as a 'fork in the road' (a '≺')—alternative trajectories splitting the future of the universe [xliv]. Reasons, values, and goals are also $\Phi$-folds within my experience, and so is my experience of a decisive reason and the subsequent decision. As $\Phi$-folds, reasons and decisions genuinely exist—they exist, physically and phenomenally, as 'forms' in the mind.

*Only what exists can cause*

On this basis, we can now ask what caused my decision. Ideally, we would employ actual causation analysis to assess 'what caused what' across two consecutive neural activity patterns in my brain, say an antecedent one when I experienced a decisive reason and a subsequent one when I decided for one of the offers. The analysis would result in a main causal account of the transition between the two activity patterns, corresponding to an actual cause-effect and an associated $\mathcal{A}$-structure.

We should also establish to what extent this causal account overlaps congruently with what exists intrinsically, because *only what exists can cause*. At one point in clock time, what exists intrinsically is a large $\Phi$-structure corresponding to my experience of a decisive reason. Soon thereafter, the state of my main complex changes, and what exists intrinsically is another large $\Phi$-structure, corresponding to my experience of having decided. To the extent that the causal account of the transition overlaps with the state of my main complex at those two points in time, we can conclude that my decision was caused by my reason, which genuinely existed and thus could genuinely cause. By the same token, since my neurons do not exist intrinsically (again, not as a mere substrate, and not as small intrinsic entities in their own right), but only extrinsically, they do not cause anything [xlv, xlvi, xlvii].

*Responsibility requires free will and self-changing actions*

While being highly conscious is necessary for free will, it is not sufficient. Free will further requires that I can envision alternatives, assess them against reasons, values, goals, and self, decide on the basis of those reasons, and intend, cause, and control actions (Tononi et al., 2022). All of that requires a substantial degree of cognitive and moral development. Only then would my main complex become competent to support the relevant $\Phi$-folds and associated cognitive processes. And only then do I become responsible for my actions, as recognized by centuries of moral and legal discernment.

Furthermore, the moment I freely engage in *self-changing* actions, I become responsible not only for what I do, but also for who I become. This is because many decisions and actions—especially life-changing decisions—will substantially change the substrate of my main complex, primarily by modifying its connectivity (Tononi et al., 2022). In principle, this could be assessed through the analysis of causal plasticity, just like the analysis of actual causation can determine what caused my action.

IIT's approach to free will and responsibility is unconventional, grounded as it is on its intrinsic ontology rather than on the issue of whether free will is compatible or incompatible with determinism. Moreover, the implications of IIT's may seem 'too good to be true' when considering free will from the extrinsic perspective. This is because from that perspective it is all too easy to *conflate existence with constitution*—'to be' with 'to be made of.' If what exists at bottom are the neurons (or atoms) that constitute my brain, and if my decision is determined 'vertically' by their state (in the sense that it 'supervenes' on them), my decision cannot be ontologically primary. Ontologically, it becomes an 'epiphenomenon.'

Furthermore, from the extrinsic perspective it is just as easy to *conflate causation with prediction*—'what causes' with 'what happens.' If my neurons (together with 'the laws of physics') are sufficient to predict what happens next, and my action is 'horizontally' determined by their mechanisms and inputs, there is no room for my decision to cause anything. Causally, I am 'carried along for the ride.'

Finally, from this perspective it is only logical to *deflect responsibility to history*—to treat one's actions as inevitable consequences of one's nature and nurture. If what each of my neurons does now can be traced back to previous occurrences along a chain of first-order (neuron-by-neuron) causations, there is no room for me to be or behave otherwise. A causal chain going back to my conception (or earlier) accounts for why my brain is constituted the way it is and thus for who I am and what I do. Even if a little indeterminism were sprinkled over the end result, that would only inject some unpredictability, rather than freedom. Accordingly, I cannot bear ultimate responsibility, not just for what I decide here and now, but also for what I have become. I am merely 'a product of my genes and environment.'

These conclusions are inevitable as long as one embraces the extrinsic perspective on what exists and causes—the assumption that what exists is a micro-physical substrate evolving over time, upon which everything else supervenes, including experiences and decisions. However, taking the intrinsic perspective changes everything. While it is perfectly fine to characterize the evolution of micro-physical substrates (more precisely, of their TPM), what matters is the unfolding of what genuinely exists and causes. A possible scenario would be that, before the evolution of large, highly adapted brains, nothing much existed intrinsically—mostly 'ontological dust.' Only then did large intrinsic entities come into being—highly conscious ones. Moreover, alternatives, values, reasons, and goals only came about when brains developed further and were refined by learning and social interactions (Tononi et al., 2022). Before that, there was no free will. Finally, the exercise of free will became the cause of further changes in connectivity (that is, of the brain's TPM), increasing one's responsibility.

As to what actually caused what, causal process analysis may trace back my action right now to a thought that occurred a few minutes ago, leading to a train of subsequent thoughts and culminating in my current decision (rather than tracing it back to my conception or earlier). Moreover, causal plasticity analysis may trace back the substrate of that thought of a few minutes ago to a pivotal free decision that occurred many years in the past, which caused enduring changes in my brain. This pivotal decision set my course for much of my life and made me responsible for my choices. Said otherwise, my conscious choices are the cause of which way things go—I can indeed steer my own future and that of our environment, rather than being carried along for the ride.

In this account, predictability is not a problem for free will. If my alternatives are evaluated based on my reasons, my reasons cause my decisions, and my decisions cause my actions, predictability should be expected. Thus, if I had strong reasons to choose the way I did, then, under the same circumstances, I should generally make the same choice. Micro-physical predictability is also not a problem. A good model may eventually offer a shortcut to predict what happens next without even bothering to unfold what genuinely exists and causes, even though that is what really matters.

In all of this, fundamental indeterminism also plays a role, but not the one that is usually assumed in arguments about free will—in short, it is necessary for existence, rather than for freedom. At the level of micro units, fundamental indeterminism (intrinsic differentiation) is a requirement for existence: to exist intrinsically, every elementary substrate unit must provide itself with a repertoire of two states of non-zero probability (differentiation) as well as specify its cause and effect state by increasing its probability (specification). This rules out both full determinism and full indeterminism. With respect to free will, the main consequence of fundamental indeterminism is that it makes the long-term unrolling of the universe, as well as of one's life, unpredictable in principle, hence not preordained. And if the future is not preordained, we are responsible not only for our actions now, but also, to the extent that our actions are self- and world-changing, for what they will bring about. While opposite to current scientific orthodoxy, these conclusions from IIT's intrinsic powers ontology and its account of causation align well with both common sense and several spiritual, religious, and wisdom traditions: we are the authors of our deliberate actions and bear responsibility for their consequences.

*Empirical tests*

How does IIT's account of free will fit with empirical evidence, and can it be tested? Neurophysiological studies regularly show some brain activity change that predicts a choice above chance before subjects become aware of their decision. Based on the dominant substrate views, this provides evidence that genuine free will is illusory: what we end up choosing is determined by the 'computations' performed by our neurons. Indeed, if we conflate existence with constitution and causation with prediction, all we will find are patterns of activity over a substrate, from which one can predict other patterns of activity, and eventually behaviors and reports.

IIT's intrinsic powers view, however, changes the perspective on what exists and causes. This is not just a matter of interpretation, but leads to testable predictions, not only about consciousness, but also specifically about free will. For instance, IIT predicts that voluntary actions, but not reflex actions, originate within the main complex in the brain. It also predicts that the $\mathcal{A}$ value and the associated $\Phi$-fold should be larger for deliberate actions vs. actions triggered by urges. Once again, propositions that may at first seem 'metaphysical'—about the nature of consciousness and free will—become fully 'physical' (in the sense of operational) and therefore empirically testable.

*Alternatives and decisions, problems and solutions, questions and answers*

There have been endless debates concerning free will, with many concluding that genuine free will is impossible, if not absurd. Remarkably, while many have doubted the existence of genuine free will, hardly anybody has expressed concern about the existence of genuine problems and solutions, or of genuine questions and answers. A likely reason is that the debate on free will has focused not on existence, but on determinism, and determinism may seem more problematic for achieving freedom than for finding answers or solutions. And yet, if we adopt substrate ontologies, we cannot have a genuine problem and solve it, or have a genuine question and answer it, just as we cannot have a genuine choice and decide how to act. In emergentist ontologies, problems and questions, like choices, may be considered emergent constructs that are necessary for explaining our behavior and are irreducible to a substrate, given multiple realizability. But just like decisions, solutions and answers must ultimately be caused by the micro-physical substrate, which is assumed to exist as such. The substrate inexorably updates its state, each state leading to the next. Once again, we are merely carried along for the ride, thinking we have solved the problem or answered the question. By contrast, in IIT's intrinsic powers framework, there are genuine problems and genuine questions, genuine solutions and genuine answers. They exist intrinsically as $\Phi$-folds within the $\Phi$-structure that corresponds to our current experience, just like alternatives and decisions. They genuinely exist, not the neurons supporting them. In fact, as already mentioned, problems and solutions, questions and answers exist *only* as contents of experience within individual subjects. They do not emerge from a neural substrate that exists as such, nor do they live in a separate, 'Platonic' realm of ideas. My neurons do not have problems, I do. And their solutions do not exist outside the minds that conceived them.

## Inferring consciousness beyond adult humans

Because consciousness is intrinsic, its presence and quality elsewhere are a matter for inference, typically guided by analogy. For other adult humans, who behave the way we behave when conscious, and have a similar body and brain, the inference is plausible enough. On the other hand, if another human is deeply asleep, behavior is insufficient to judge the presence and quality of experience: the person may just as well be conscious (dreaming sleep) or unconscious (dreamless sleep). We may still obtain a plausible answer by awakening the person and trusting their retrospective report, just as we would trust ours. Scientific discoveries can further inform inferences based on analogy. For example, having established that low levels of EEG slow wave activity in posterior-central cortical regions are reliably associated with dreaming sleep, we can be confident that the sleeper is conscious without awakening them. We may even be able to infer some of the contents of their dreams based on the location of high frequency activity or by employing sophisticated decoding techniques.

However, when it comes to pathological conditions, to early development, to species clearly different from us, not to mention to artifacts built in a way that is radically different, analogy becomes a less reliable guide—in fact, it may become utterly unreliable. Below, we briefly consider some examples in each category, primarily to illustrate how, in such cases, one might draw inferences based on IIT's principled approach. Given the many assumptions required, such inferences should be treated with caution, yet they remain preferable to guesses based on 'gut feelings.'

*Patients*

There are many conditions in which it is hard to determine the presence of consciousness even in adult human beings. For example, it is not easy to know whether patients with parasomnias, such as sleep-walking and sleep-talking, were conscious, or partially so, during a particular episode. Retrospective reports can be used but they are harder to trust than, say, awakening from natural sleep. Similar difficulties occur with epileptic subjects with complex partial seizures who appear confused and engage in automatisms, as well as with subjects with absences and other generalized seizures. Anesthetics and, more generally, drugs inducing profoundly altered states, such as ketamine or 5-MeO-DMT, pose similar interpretive problems. In some cases, the rapid and careful collection of retrospective reports can be highly suggestive, especially if coupled with reliable neurophysiological indicators of consciousness (Sarasso et al., 2015). In other cases, one may have to rely exclusively on such indicators.

The evolving clinical practice with patients with disorders of consciousness is beginning to reflect these developments. A breakthrough was the demonstration, using neuroimaging, that a behaviorally unresponsive patient could activate the same brain areas as healthy subjects when instructed to imagine different scenarios (playing tennis or navigating her room) (Owen et al., 2006). Unfortunately, even subjects who are behaviorally conscious but neurologically impaired may fail such cognitively demanding, 'active' paradigms (Monti et al., 2010). Neural correlates of cognitive functions typically engaged when we are conscious can also be employed. However, a failure to show cognitive activations should not be taken as evidence of unconsciousness (Boly et al., 2007). In fact, there are strong reasons to infer that brain-damaged patients with minimal responsiveness—who are barely capable of tracking faces or objects with their eyes—have rich sensory experiences (Cecconi et al., submitted). One reason is that, in such patients, PCI (mentioned above) is almost always in the range of overtly conscious subjects. PCI, a perturbational method based on TMS-EEG that has been extensively validated in many conditions of consciousness and unconsciousness in healthy adults, offers the most sensitive and specific assessment of the presence or absence of consciousness. Crucially, the evaluation of PCI does not just have empirical validity, but also construct validity, having been developed based on IIT's postulates to capture some of the essential properties of consciousness (albeit crudely).

The broader point is that the determination of consciousness and its quality (for instance, the presence of pain) should ultimately be based on a theory of what consciousness is, and what its substrate must be like to support it, rather than on purely empirical correlates. Otherwise, how would one hope to determine whether catatonic subjects are conscious or not, especially if they have amnesia for the episode? Or subjects with terminal dementia? Or patients who may not show signs of cognitive activations, yet show differentiated neural activity when presented with movies or speech? Or patients in which only a few 'brain islands' seem anatomically and metabolically preserved, albeit at much lower levels? What should we make of such patients? Do they have a fraction of the quantity of consciousness of healthy subjects? Do they just hear sounds or feel pain, depending on which cortical island is spared? In other words, 'what is it like' to be a brain island, if it feels like anything at all? And how big must the island be to qualify? In such cases, reasonable inferences can only be based on extrapolations grounded on a theory that has been validated in ourselves under normal conditions, one that accounts for both the quantity and quality of consciousness.

*Infants*

Similar riddles occur when considering the question of consciousness in human development. What is it like to be a newborn baby with an immature brain and restricted connectivity within and among brain regions? Some think that infants may be very much like adults with respect to basic aspects of experience. Others have proposed that infants may be even more conscious than adults because, unlike us, they tend to attend to everything at once. Yet others suggest that consciousness as we know it may only appear with language, so infants would be essentially eating, smiling and crying machines until they start babbling.

Thankfully, much progress has been made in characterizing the development of brain and behavior before and after birth (Padilla & Lagercrantz, 2020). This has led to a more refined understanding of the factors that, largely based on analogy, might indicate how consciousness emerges during ontogeny, whether progressively or in bouts, whether as a whole or as modality-specific domains that fuse later on (Padilla & Lagercrantz, 2020). Moreover, various neurophysiological markers previously employed in adults are being evaluated in preterm infants, newborns, and during the first months (Passos-Ferreira, 2024).

Can IIT help shed some light on consciousness in human development? At present, what IIT can do is to provide, rather than a tentative answer, some theoretically motivated guidelines. For example, work in rodents indicates that, during early development, neurons within local thalamo-cortical modules may already be active and connected to sensory inputs or motor outputs, but not yet among themselves (reviewed in (Martini et al., 2021)). Based on IIT, we can infer that no complex of high $\Phi$ can be present in the cortico-thalamic system, owing to a complete lack of integration. Later on, neurons within primary sensory areas go through an explosive growth of connections, but initially these connections are arranged in a rather random manner (Martini et al., 2021). In this case, it is the lack of differentiation, rather than of integration, that severely limits the $\Phi$ value achievable by a complex of neurons in primary areas. Still later, after a process of synaptic refinement, the connectivity matures to achieve an adult-like pattern, which in sensory areas resembles a lattice. Suddenly, the neural substrate may become capable of specifying $\Phi$-structures characterized by extension, with much higher values of $\Phi$. Further refinement and synaptic changes due to learning will lead to the formation of local cliques; to the maturation of intra-areal connectivity in cortical regions higher up in the sensory hierarchy; and to the establishment of adult-like pattern of inter-areal connectivity, configuring multiple hierarchies that will substantially expand the main complex. To the extent that similar developmental changes occur in the human nervous system, we would expect parallel changes in the quantity and quality of consciousness.

*Other species*

With healthy, adult humans, the so-called 'problem of other minds' is mostly academic. Yet it is anything but when we consider species substantially different from us. Is there anything it is like to be them, and what would it be like? Intuitions have varied greatly, influenced by both historical and cultural factors. Naturally enough, analogy can serve as a good first step. Most people would find it hard to deny consciousness to other primates, given their remarkable similarities to us.

The continuity in behavior, body, and brain that ties us to other mammals, including those we think we understand better, like dogs and cats, suggests that they, too, may be 'sentient.' Again, analogy becomes more dubious with species radically different from us. However, careful studies of animal behavior have invariably revealed great 'intelligence' and 'sophistication,' including 'emotional' intelligence and 'cognitive' sophistication. Because we ourselves only show these traits when we are conscious, it is natural to extrapolate that they may be accompanied by 'sentience' elsewhere in nature. Visual recognition of self, meta-cognition (knowing one's mind), theory of mind (making assumptions about the mind of others), empathy and long-range planning have all been demonstrated in primates, rodents and other orders. Birds such as ravens, crows, magpies, and parrots, fish such as tuna, cephalopods such as octopuses, and insects such as bees can also perform sophisticated, non-stereotyped behaviors, as well as learned behaviors that we associate with consciousness if carried out by people.

When it comes to the brain, there are many more similarities than differences between ours and that of other mammals. Cell types, synapses, proteins, and genes in the human brain are also similar to those found in many other species. The brains of elephants, dolphins, and whales surpass ours in size and number of neurons. Even brains very different from ours can be extraordinarily complex. For example, a bee has around 800,000 nerve cells of remarkable diversity, packed together much more densely than in our brain and forming complicated networks and circuits.

However, while complex brains and behaviors are suggestive of consciousness, they do not guarantee it. As already discussed, the cerebellum is by most measures an extraordinarily rich and complex brain, heavily interconnected with the cerebral cortex, yet it does not seem to contribute any content to experience. Is the bee's brain central complex more like the cerebellum with respect to experience, or is it more like the cerebral cortex? Also, we can perform some complex behaviors in a seemingly automatic manner, unaccompanied by experiential content. Does a been decide whether to turn right or left in a maze as we would do when we deliberate consciously, or as we do when we use the right or left hand to push a key on the keyboard when typing?

In the end, the quantity and quality of consciousness in other species is again best addressed as an inference from a good explanation of our own consciousness. In this respect, perhaps the most important inferential basis should come from detailed connectomes, which begun with that of a worm and are now expanding to other organisms, such as the fruit fly. Fine-grained neuroanatomy, complemented by the relevant neurophysiology, should permit some initial conjectures about the presence of a large main complex. Thus, if it turned out that a bee brain contains a main complex with an extrapolated $\Phi$ value that is only an order of magnitude less than that of a human brain during dreaming sleep, and much higher than a human brain in dreamless sleep, we would be justified in assuming that there is in fact something it is like to be a bee. An analysis of the kinds of $\Phi$-folds that can be supported by a bee's brain could also suggest whether bees can experience, for example, space, time, and pain.

*Machines, organoids, and the double dissociation between consciousness and intelligence*

We have been living together with other animals for a long time and, when their behavior show evidence of what we consider intelligence, empathy, or suffering, many of us have assumed that they may be conscious, though likely in different or lesser ways. But only now are we confronted with machines that we have built ourselves and whose intelligence may soon equal or surpass our own, by any definition. Whether they may or may not be conscious, either now or in the near future, is controversial.

During the past decades, it was often assumed that machines could not possibly be conscious because of some critical deficiency compared to us, whether the ability to reason, to learn, to imagine, to be creative, to show empathy or a sense of humor, to produce art, music, or literature, and so on. Now however, with machines regularly passing the Turing test (being indistinguishable from humans in conversation), and with the impending arrival of artificial general intelligence (AGI), many are suddenly less sure. Once machines become functionally equal to us or better than us, on what grounds should we deny them the prerogative of consciousness? In fact, some have readily accepted the possibility that highly intelligent machines—artificial brain with or without an artificial body—may indeed be conscious. After all, the reasoning often goes, we don't really know what consciousness is, so we might as well err on the side of caution. Machine consciousness is also in line with the dominant computational-functionalist paradigm, which assumes that consciousness, if it is anything at all, must correspond to some high-level function(s) (though it is not yet clear which one) (Butlin et al., 2023). Therefore, once machines equal us in *every* function, they would necessarily be conscious. If somebody argues that machines are in a sense just immense look-up tables, then the sobering conclusion might as well be that we, too, are ultimately giant neuronal look-up tables, which came about by trial and error during evolution.

However, some are beginning to voice a degree of skepticism, perhaps because AGI is becoming a reality, rather than a far-fetched possibility. Furthermore, precisely because we know all there is to know about the machines we built, it seems even harder to imagine how 'subjectivity' would emerge out of such look-up tables. This is leading some to suspect that consciousness may require some special ingredient provided by biology. Once again, however, what that ingredient would be, and why it would account for the presence and quality of experience, remains unexplained.

IIT's approach to the issue of machine consciousness is the same as with consciousness in disease, ontogeny, and phylogeny—namely inference from a good explanation, where the explanation is provided by its consciousness-fist paradigm, its operationalization, and its validation in ourselves.

In fact, in the case of machines, the inferences one can draw from IIT are already clear. It can be shown that, if the machine's brain is implemented in a von Neumann computer (whether embodied in a robot or not is irrelevant), then it may be functionally equivalent to us without being phenomenally equivalent in the least. More precisely, it can be demonstrated that a simple von Neumann–style computer that simulates an even simpler system of logic gates can be functionally equivalent to it without specifying an equivalent cause–effect structure. In fact, the computer 'disintegrates' into many small modules, each of which supports trivially small $\Phi$-structures, which remain largely the same regardless of what system is being simulated. This result holds for computers of arbitrary size (Findlay et al., 2024). Therefore, if IIT is right, computers that may replicate our behaviors or cognitive functions will not replicate our experiences.[xlviii] Thus, returning to the contrast between functional and phenomenal vision, a computer vision system in a self-driving car would act as if it did 'see'

*functionally*—recognize scenes and objects and move around the world much like we would—yet not see anything *phenomenally*. In other words, it would not genuinely exist—there would be nothing it is like for it to exist [xlix].

Altogether, IIT implies that function and phenomenology, intelligence and consciousness, can be doubly dissociated. Not only can we have instances of high intelligence without consciousness, as in the case of AI, but there are clear cases of rich experiences unaccompanied by intelligent behavior. For example, as already mentioned, states of pure presence experience demonstrate that there consciousness can be intensely present and vivid while being completely dissociated from any ability to pursue goal-directed behaviors—in fact, experience is devoid of any thought, sense of self, and perceptual objects (Boly et al., 2024). Similar states of pure experience can also occur under the influence of certain psychedelic drugs. More commonly, many dreams demonstrate that we can be vividly conscious while lacking the ability to reflect, remember, and report (Tononi et al., 2024). Also, raw experience can momentarily overwhelm cognitive functions during intense pain or sudden loud stimuli. Furthermore, as discussed in the section on the richness of consciousness, the content of every experience can only feel the way it does by being composed of an exceedingly large number of phenomenal distinctions and relations, yet most of this content cannot be remembered, reported, or put to functional use. As was also mentioned, patients who are minimally interactive owing to severe brain damage, especially in prefrontal regions, are likely to harbor rich sensory experiences despite being cognitively impaired (Cecconi et al., submitted). More speculatively, we may soon be faced with organoids engineered to reproduce the internal organization of our posterior-central cortex. Just like our dreaming brains, such organoids may become highly conscious, presumably in a predominantly spatial manner, regardless of whether they are connected to the environment via sensors and effectors, and regardless of whether they can perform any 'intelligent' or 'cognitive' function.

In conclusion, we should be mindful that consciousness may be present where we may not expect it—say, in the absence of intelligence—and careful in attributing it to anything that appears complex or intelligent, including AI. In every instance, according to IIT, the presence of consciousness and its quality depends not on intelligence, but on the presence and internal organization of a large main complex.

On the other hand, IIT provides some rationale as to why selective pressure for intelligence may have favored the emergence, during evolution, of neural networks capable of supporting high levels of consciousness. Simulations using simple 'animats' evolving in a 'rich' environment show the preferential emergence of integrated substrates, rather than modular ones (Albantakis et al., 2014). The explanation is that, for the same number of units, an integrated substrate can pack more functions, and this can be evolutionarily advantageous given constraints on energy consumption, wiring, and time. One should also consider that the 'spatial' smoothness of the environment and that of sensors, as well as neurodevelopmental processes, may favor the emergence of grid-like substrates. In these and other ways, selective pressure for functional prowess may carry along phenomenal richness [l].

## Some instructive criticisms of IIT

Before concluding, it may be helpful to briefly consider some criticisms of IIT. The goal is not to provide a thorough reply, but rather to illustrate how objections or misunderstandings of IIT often stem, directly or indirectly, from an incomplete appreciation of its consciousness-first starting point and its development of an intrinsic ontology.

*Information as observer-dependent*

A paradigmatic case is that of John Searle, who wrote a long critique of IIT based on his valid argument that consciousness is ontologically subjective and observer-independent, whereas 'information' depends on an external observer (Searle, 2013). As should be clear from this and previous expositions, IIT fully agrees with Searle: experience is ontologically subjective, that is to say, intrinsic. Moreover, much of IIT's development has focused on defining information from the intrinsic perspective, in contrast to Shannon information (see *Box 1*) (A. Zaeemzadeh & G Tononi, 2024). It is as if Searle's interpretation of the word 'information' led him to read the entire theory as saying the opposite of what it says. Perhaps for the same reason, a similar misunderstanding of IIT appears in a recent book whose very title and theme, 'The Blind Spot,' rightly emphasizes the importance of starting from phenomenology (Frank et al., 2024), [li].

*Functionalism*

Another example of misreading involves various interpretations of IIT as 'functionalist.' IIT has consistently highlighted the radical difference between its intrinsic, consciousness-first paradigm, and the extrinsic, computational-functionalist paradigm that dominates psychology, cognitive neuroscience, and computer science. That IIT implies a double dissociation between consciousness and intelligence has also been repeatedly emphasized. Yet despite this emphasis, IIT has sometimes been labelled a theory of intelligence rather than of consciousness, and listed among functionalist approaches (Block, 2009). As already mentioned, functionalism identifies a mental state with what it does—a function or causal role—rather than with what it is. By contrast, IIT characterizes an experience by what it is—a cause-effect structure—rather than by what it does. Viewing IIT as functionalist seems to reflect an extrinsic perspective so engrained that alternative interpretations may be overlooked even when explicitly presented.

*The 'unfolding' argument*

Along similar lines, consider the so-called 'unfolding argument' (Doerig et al., 2019). This appeals to the proof that feedforward systems can be functionally equivalent to recurrent ones (Hornik et al., 1989), a proof originally invoked by IIT to argue that functional equivalence does not entail phenomenal equivalence (Oizumi et al., 2014). Ironically, the unfolding argument invokes the same proof to argue that 'causal theories' of consciousness (namely IIT) are either wrong or untestable. The reasoning is that there would be no way to conclude that a system may be conscious (the recurrent one) and the other one not (the feedforward one) if their input-output behaviors/functions are the same. This argument serves as a revealing example of how one's presuppositions (in this case, functionalist ones) can make it difficult to even entertain IIT's intrinsic perspective. Indeed, if consciousness is defined in functional terms—by what it does, as assessed by behavior—then by definition functionally equivalent systems become consciousness-equivalent, regardless of how they are organized internally. But if consciousness is defined phenomenally—by what it is, as assessed by introspection, as it should— then its properties can only be accounted by substrates with a suitable internal organization, regardless of their input-output behaviors/functions.

*Grids*

Another interesting case is that of Scott Aaronson who, after demonstrating that IIT implies the possibility that large grids could be highly conscious, concluded that the theory was therefore untenable (while appreciating its mathematical precision). (Aaronson, 2014). [lii]. At issue is not Aaronson's demonstration that large grids could have high $\Phi$, something that IIT had already pointed out (Balduzzi & Tononi, 2008). Instead, it is Aaronson's intuition that grids cannot possibly be conscious, seemingly because they are too simple. Yet intuitions, even those widely shared, may not be the best guide in deciding whether something is conscious or not (much as intuitions may not be the best guide in deciding whether the earth moves or not). As already emphasized, any inference about the status of grids (or any other substrate) can only be made based on a theory like IIT—one that characterizes consciousness phenomenally, establishes what substrate can support it, and can be validated empirically in us. Interestingly, when it comes to grids, the evidence strongly aligns with IIT (Haun & Tononi, 2019; Tononi, 2014). As we have seen, most of our experiences feel spatially extended, which implies a rich phenomenal structure composed of spots bound according to reflexivity, inclusion, connection, and fusion. Furthermore, only grids with dense connectivity, properly arranged, can specify a cause-effect structure that is equally rich and accounts for the feeling of spatial extendedness. As it turns out, most of our posterior-central cortex is organized like a hierarchy of grids—in other words, much of the substrate of consciousness is grid-like. And finally, manipulating these grids can alter or abolish corresponding aspects of experience, including the overall experience of space. In summary, whether we find the notion of grid-based substrates intuitive or not, both phenomenology and empirical evidence indicate that grids in the brain deserve serious consideration as a substrate for spatial experience.

*Emergence and dualism*

Presentations of IIT often begin by stating explicitly that we cannot assume 'physics' as primary and hope to 'squeeze' consciousness out of it. Instead, we must start from consciousness, whose existence is immediate and irrefutable, and try to account for it in physical terms, where 'physical' is meant operationally (taking and making a difference). Yet by the end of the presentation, a common reaction is to express appreciation for IIT's theoretical foundations, only to ask: 'but why would a cause-effect structure *give rise* to experience?' Apparently, despite IIT's unrelenting emphasis on the primacy of experience, it is hard to overcome the inclination to put the 'physical' first and treat the mental as 'emerging' from it. Just as often, IIT's proposed identity between experiences and cause-effect structures is understood as implying a dualism—a correspondence between two different things, and one that remains to be explained. Yet IIT makes it clear that cause-effect structures are meant precisely as an *explanation* of phenomenal properties in terms of physical properties (understood operationally), not as something that is correlated with them.

*Panpsychism*

Another common criticism of IIT is that it would open the way to panpsychism—the idea that consciousness is ubiquitous in nature. For many, this seems to imply an automatic disqualification from science. On the other hand, if one takes the existence of consciousness seriously, some kind of panpsychism may be a more parsimonious starting point than both dualism and physicalism, the latter seemingly requiring the 'brute emergence' of experience out of non-experience (Strawson & Freeman, 2006). However, panpsychism has little to offer to explain what we know about the substrate of consciousness, and no account of how our experiences would come about from the composition of elementary experiences (the so-called combination problem). But what does IIT actually imply about the presence of consciousness in nature? Assuming that IIT is sufficiently validated in us, we can infer that consciousness should not be attributed indiscriminately to all things—far from it. The exclusion postulate implies that there can be no superposition of consciousnesses and no aggregate consciousness. As discussed in an earlier section, if the main complex is constituted by a set of neuronal groups in posterior-central cortex, then no subset of it, superset, or paraset of it, at any grain, can support other consciousnesses. And given that each of us is conscious, neither can groups of people or societies. Furthermore, while IIT implies that consciousness can be graded, it also implies that the differences in consciousness can be extreme. As already mentioned, the quantity of consciousness $\Phi$—the sum of the $\varphi$ values of the distinctions and relations composing a complex's $\Phi$-structure—can grow double-exponentially with the number of intrinsic units of the complex. When we are conscious, the main complex in our brain should support $\Phi$-structures having literally hyper-astronomical values of $\Phi$. On the other hand, during dreamless sleep, the main complex would disintegrate into a multitude of mini-complexes, each of which would support a cause–effect structures of minimal $\Phi$. Accordingly, each mini complex would feel like hardly anything at all, in agreement with our sense that we must have 'lost consciousness.' As a metaphor, a temperature of -270 ºC, while not absolute zero, is certainly not 'warm' in any meaningful sense. Similarly, mini complexes should not be considered conscious in any meaningful way. In summary, inferences based on IIT would currently support a view where most things, whether subatomic particles, atoms, molecules, stones, trees, bodies, continents, planets, or galaxies, are either aggregates of smaller complexes or have negligible $\Phi$. They would then not be conscious at all, or not meaningfully so (Tononi et al., 2025).

## IIT and the place of consciousness in nature

The success of science has brought with it some beliefs that are often accepted as true. A foundational one, since the times of Galileo, is that science should take the *extrinsic perspective*: to do science properly, we must remove the subject and develop an ontology—a notion of what exists—that is fully objective. At its heart, the Galilean stance assumes that what exists is a universe of matter and energy, and that all natural phenomena should be accounted for on this basis. This approach has been immensely fruitful and has changed not only the world around us, but also the current scientific understanding of our place in nature, leading to several 'displacements.' We have learned that, rather than looming large at the center of the universe, we are mere specks of matter relegated to a far corner (the *cosmic displacement*). Rather than having an essential soul, we are ultimately biological machines—bags of atoms behaving in complex ways (the *neurobiological displacement*). As Francis Crick put it, " 'You,' your joys and your sorrows, your memories and your ambitions, your sense of personal identity and free will, are in fact no more than the behavior of a vast assembly of nerve cells and their associated molecules" (Crick, 1994). Rather than the pinnacle of creation, we are the product of chance and blind selection (the *evolutionary displacement*). And we are rapidly moving toward a fourth displacement: not only are we biological machines, but we may soon be sidelined by artificial machines—machines that can do all we do, only better, faster, and more reliably (the *AI displacement*).

In short, according to the dominant scientific worldview, which takes the extrinsic perspective on what exists, this is how things really are: our place in the universe is both peripheral and ephemeral, all we feel and think is ultimately a byproduct of neural computations, our origin is an evolutionary game of chance and circumstance, and our inner life may soon be eclipsed by AI. Inevitably, such a perspective can underlie a sense of meaninglessness. If consciousness is merely 'carried along for the ride' by a machine that 'runs through the motions,' then where is the meaning of what we see, hear, feel, think, and strive for—which is all that exists for us? This is not just a cognitive dissonance between what we feel and what we know. It is a crisis—and one that is indeed 'existential,' as it hinges on our vision of what exists.

As we have argued, however, the extrinsic perspective cannot account for what consciousness is, and cannot lead to a complete scientific worldview, because it inverts the natural order of things. IIT's intrinsic ontology, instead, takes experience as primary, because it exists immediately and irrefutably. Everything else, including the extrinsic existence of what we call matter and energy, is an inference from within experience. Doing so can hopefully lead to a systematic account of what consciousness is in a way that fits what we know about the brain. It can also open the way to a unified view of nature. From the intrinsic perspective of human experience, we can still properly characterize the universe around us in operational terms. But we can also hope to understand the universe within us, encompassing feelings, meanings, values, goals, freedom, and agency.

In fact, IIT's intrinsic ontology reverses the four displacements mentioned above. If IIT is on the right track, human consciousness is far from being a minuscule flicker among giant stars and galaxies; more likely, measured by intrinsic existence ($\Phi$-structures and their $\Phi$ value), it is the largest kind of entity we know. Far from being 'carried along for the ride' by our neurons, we exist, and not them. We—not our neurons—have meanings, reasons, values, and purposes. Those only exist in our mind, as do all the concepts and all the sciences, as does all of art, music, and literature. And we have free will and responsibility—nothing else does. Far from being the mere product of chance and natural selection, through self-changing actions we are very much the makers of who we become. Finally, far from being superseded by AI, we are beings that exist intrinsically, for ourselves, while super-intelligent machines are merely tools, which only exist for us and which, unlike us, are mere aggregates of ontological dust.

---

[i] We use the qualifier ‘genuine’ for lack of a better word, to emphasize that experience, i.e. ‘intrinsic existence,’ is existence in the only sense intrinsically worth being, while realizing full well that ‘extrinsic existence’ (see below) is not ‘spurious’ or ‘fake’.

[ii] The expression ‘what it is like to be’ was introduced by Farrell (Farrell, 1950) and Sprigge (Sprigge, 1971) and made popular as a definition of consciousness by Nagel(Nagel, 1974). The 0th axiom also resembles Descartes’ ‘I think therefore I am,’ although for Descartes this truth represented the starting point for epistemology, whereas for IIT it is the starting point for ontology.

[iii] Another way to express the axioms is to say that, to genuinely exist, an entity must satisfy *self-ness* (being for itself), *this-ness* (being this one), *one-ness* (being one), *all-ness* (being all it is), and *thus-ness* (being the way it is)—the five essential properties of being.

[iv] IIT takes the set of axioms to be complete: there are no further properties of consciousness that are true of every conceivable experience. Some properties, like an experience being sweet or sour, are obviously non-essential, that is, they are accidental. But even properties that may seem candidates for axiomatic status fail the conceivability test. These include space (experience typically takes place in some spatial frame), time (an experience usually feels like it flows from a past to a future), change (an experience usually transitions or flows into another), subject–object distinction (an experience seems to involve both a subject and an object), intentionality (experiences usually refer to something in the world, or at least to something other than the subject), a sense of self (many experiences include a reference to one’s body or even to one’s autobiographical self), reflectiveness (experiences often include awareness that one is experiencing whatever one is experiencing), figure–ground segregation (an experience usually includes some object and some background), situatedness (an experience is often bound to a time and a place), will (experience offers the opportunity for action), and affect (experience is often colored by some mood). Yet for all these properties, it is conceivable that an experience might lack them. For example, we can conceive of an experience that lacks spatial aspects, say, a purely olfactory experience; of an experience without time, as in certain meditative states or through the effect of some psychedelic drugs; of one that stays the same without changing; of one without the sense of subject–object distinction; of one without an intentional object (say, boredom); of one lacking a sense of self, as in some meditative practices or even when absorbed in an action movie; of one that is non-reflective, like seeing the screen without thinking about it, and so on.

[v] Note that negating the postulates, as opposed to the axioms, does not lead to absurdity. This is because the postulates are extrinsic/objective, and from the extrinsic perspective it is perfectly possible to consider of cause-effect power exerted by something on something else, rather than on itself; or to consider an average effect, rather than a specific one; or to consider a system that is not integrated; or one with arbitrary borders and grain; or to ignore structure.

[vi] It should be understood that axioms, postulates, principles, and assumptions of IIT do not arise in a vacuum but are the result of an attempt to arrive at a ‘good explanation’ of the facts about consciousness by a conscious being capable of introspection and reasoning, relying especially on the principle of sufficient reason.

[vii] Of course, such complete description is only possible for idealized systems. For realistic systems, the analysis of cause-effect power must rely on several assumptions and resort to heuristics and approximations.

[viii] Although the Eleatic principle says ‘take *or* make a difference.’ Moreover, it does not emphasize the necessity of ‘providing’ a difference to begin with, upon which a difference can be taken and made.

[ix] The principle of becoming governs how the TPM evolves at every update of the universal substrate depending on what happens (i.e. which states actually occur changes the probabilities in the TPM, which in turn influence which states will occur).

[x] Expressed with respect to intrinsic entities over a substrate, the *principle of maximal existence* (among candidate intrinsic entities, the one that actually exists is the one that would exist the most), combined with the *principle of sufficient reason* (there must be a reason why an intrinsic entity should be the one it is and not another) and with the *principle of identity of indiscernibles* (if two candidate entities have all the same properties, then they are one and the same intrinsic entity), imply the *principle of no free (intrinsic) existence* (a substrate unit can only support the cause-effect power of a single intrinsic entity). Together, these principles also imply that a substrate unit can only support the intrinsic cause-effect power of a single intrinsic unit (and cannot mediate that of other intrinsic units), and that a background unit can only mediate the intrinsic cause-effect power of a single intrinsic unit. The maximal existence principle and the no free existence principle in IIT’s ontology are somewhat analogous to the least action principle and conservation principles in physics, respectively (conservation principles are consequences of least action given some symmetries).

[xi] *Intrinsic differentiation* captures the notion that a system must provide a repertoire of different states with non-zero probability over which it can take and make a difference (existence), to do so by itself (intrinsicality), and to do so in its current state (specificity) (Mayner et al., 2025). *Intrinsic specification* captures the notion that a system must be able to take and make a difference by increasing the probability of a specific cause-effect state from its repertoire (existence), to do so upon itself (intrinsicality), and to do so in its current state (specificity). Intrinsic differentiation is assessed by the intrinsic difference from maximal specification (only one state available, with probability 1). Intrinsic specification is assessed by the intrinsic difference of a state from maximal differentiation (all states available, with uniform probability). By IIT’s principle of minimal existence, a system cannot exist more than the requirement of existence it satisfies the least, i.e. more than the minimum of intrinsic differentiation and specification for any candidate state. By the principle of maximal existence, the state differentiated and specified by the system is the one that maximizes this minimum—i.e. the one that maximizes *intrinsic information*(Mayner et al., 2025). Note that intrinsic differentiation, as defined above, is a property of a system state, whereas notions of phenomenal differentiation, perceptual differentiation, and neurophysiological differentiation (see below) refer to system averages.

[xii] It should be possible to demonstrate that $\varphi_s^*$ uniquely satisfies the postulates of integration and exclusion. Moreover, it should be possible to demonstrate that the postulates, complemented by IIT’s principles, are necessary and sufficient to uniquely specify an entity and its cause-effect power.

[xiii] Increased selectivity can arise at macro grains because the analysis of cause-effect power treats each macro state of the macro units as equally likely, corresponding to a non-uniform distribution of micro states.

[xiv] Metaphorically, we can unfold a full cause-effect structure from a neural network (which is convenient for manipulation and observation), just as we can unfold a tent from its packed substrate (convenient for transportation), an organism from a string of DNA (convenient for reproduction) or, more abstractly, a data file in the original format from its zipped version (convenient for communication).

[xv] Incongruent distinctions are not components of the complex and its specific cause–effect power because they would violate the information postulate, according to which the experience can only be ‘this one.’

[xvi] ‘Supports’ is used here for lack of a better word but without any implication that the substrate would exist ‘as such,’ like the foundations of a house, and that the experience would be built upon it by adding extra ingredients. What exists, according to IIT, is the substrate unfolded.

[xvii] Cortical minicolumns, with a diameter of the order of tens of microns, may have high intrinsic information and be highly irreducible owing to the extremely dense, recurrent local connectivity. Most connections among pyramidal neurons occur within 200-300 microns, and they are especially dense in supragranular layers.

[xviii] It should be noted that a complex cannot specify intrinsically (i.e. through its own mechanisms) all the distinct concepts that could be specified over its states.(Feldman, 2003) Therefore, in a relevant sense, no system can fully conceptualize itself.

[xix] More precisely, substrates constituted of units with special input/output properties, organized like rooted trees, and embedded in pyramids of grids, can specify structural motifs such as concepts bound to configurations of features, as well as extensions thereof, such as hierarchical inclusion, connection, and fusion.

[xx] Another possibility would be to expose a neural network simulating primary or secondary cortices to the local statistics of natural images as filtered by sensory organs (for example, retinal cones). The resulting connectivity might provide a good enough substrate to unfold kernels similar to those found in those areas.

[xxi] The identity of quality and structure is incompatible with ‘inverted spectrum’ arguments.

[xxii] In IIT, a structure, including that of a kernel, is causal and intrinsic, rather than a projection of variables and relations among them onto an extrinsically defined space.

[xxiii] Consistent with pain corresponding to a compound content, patients with pain asymbolia are able to discriminate 'painful' stimuli from other sensory stimuli, but they do not experience the affective component, as if the 'painfulness' of pain were missing.

[xxiv] In essence, the meaning in a box thought experiment assumes that there is a (unique) mapping from a detailed causal model of a system in a state, as expressed by its TPM, and the presence and quality of the experience it supports. IIT provides this mapping based on the essential properties of consciousness (axioms) formulated operationally (postulates) so that they can be applied to the TPM.

As an even more bare-bones example, consider the experience of a central red spot over a dark background (again, say, as the last scene of a dream). The neural correlate of such an experience might be the activation of a few neuronal groups (minicolumns) in primary and secondary visual cortex, for a fraction of a second, with hardly any activity elsewhere in the corticothalamic system. How can the activation of those minicolumns mean (and feel like) 'a central red spot over a dark background'? According to IIT, the lattice-like, hierarchical intrinsic connectivity among minicolumns in posterior-central cortex unfolds into an immense cause-effect structure that accounts for the feeling of spatial extendedness and for hierarchical conceptual structures. This applies even if most minicolumns that constitute the lattice are OFF, as long as they form a complex. The feeling of a spot 'located' in the center of the field would be accounted for by a 'warping' of distinctions and relations specified by neuronal minicolumns at that spot and their relations of inclusion with respect to other spots. Finally, the 'red' quality of the spot would be accounted for by the specific local subs-structure (kernel) supported by the local clique-like intrinsic connectivity among minicolumns typical of visual cortex (while cliques elsewhere in the visual cortex unfold as 'black' kernels). This entire structure is needed to account for the meaning/feeling of the experience.

[xxv] An intriguing conjecture, inspired by IIT, is that alongside the rapid dynamics of neural activity triggered by a stimulus, there may occur equally rapid changes in the strength of neuronal interactions, especially within the dense lattice of posterior-central cortex (Mayner et al., 2024). Such changes may facilitate a 'compositional search' that allows mechanisms of various orders—from single neurons to the entire main complex—to satisfy at once local constraints, which can be settled rapidly, and global constraints, which can help to establish a congruent activity pattern. Most important, such changes can 'endorse' the activity pattern triggered by a stimulus. By multiplying the strength of its intrinsic connectivity in a state-dependent manner, the main complex can specify the very same activity pattern that was triggered extrinsically, but do so intrinsically, with maximal cause-effect power, and over its macro time scale.

[xxvi] It is tempting to consider the role of the stimulus as a trigger as somewhat similar to the role of a 'prompt' in generative models.

[xxvii] It has a much larger number of units and can go through a much larger number of states.

[xxviii] The repeated alternation between periods of synaptic up-selection during wakefulness, when the brain is connected to the environment, and synaptic down-selection during sleep, when it is disconnected but spontaneously active, is likely to play a critical role in promoting matching (Tononi & Cirelli, 2014).

[xxix] For example, it can be shown that the sum of integrated information of relations is maximized when the distinctions' integrated information values are proportional to the size of their purviews, which is not attainable in systems with random connectivity profiles (A. Zaeemzadeh & G. Tononi, 2024).

[xxx] It should be emphasized that IIT's research program in no way dismisses the importance of characterizing cognitive functions and investigating the underlying neural processes. That is the bread and butter of cognitive neuroscience, and countless studies are dedicated to that goal. However, several mainstream approaches to consciousness, such as global workspace theory (Mashour et al., 2020) and higher order theories (Brown et al., 2019), consider consciousness, which they revealingly call 'conscious processing,' as another cognitive function, such as the detection and broadcasting of information, or the monitoring of incoming information. By contrast, IIT's consciousness-first approach sees experience as ontologically primary, and cognitive processes as auxiliary functions. In other words, once consciousness and its neural substrate are properly understood, they should be at the center of a functional understanding or the brain, rather than a baffling epiphenomenon of what the brain does.

[xxxi] The information-processing approach common in psychology and neuroscience estimates the information bandwidth of human consciousness to be at around $7 \pm 2$ items or $\leq 40$ bits per second, which may seem paradoxical given that sensory transmission rates are many orders of magnitude larger (for a discussion, see (Tononi et al., 2016)). As argued above, this approach ignores or denies the richness of experience, which is not measured by Shannon information, but by the integrated information associated with a $\Phi$-structure, whose amount is bound to be extremely large (see *Box 1*). Even so, we can expect the bandwidth of the triggering of specific $\Phi$-structures by sensory inputs, which are highly parallel, to be large. By contrast, we can expect the bandwidth of attention, working memory, executive function, and ultimately behavioral output to be low, being mediated by processing loops that operate serially. Such processing loops are highly flexible and allow us to access and report the state of disparate subsets of the main complex, but only a few at a time. More fundamentally, regardless of channel capacity, we have no way to 'dump' or transmit the rich $\Phi$-structure that composes every fleeting experience—we can only hope to trigger a similar one in subjects with similar main complexes by exposing them to similar sensory inputs, or by communicating through language.

[xxxii] In other words, there cannot be a difference in high-level properties without a difference in low-level ones.

[xxxiii] Not reducible here means that a high-level function cannot be identical to its low level, microphysical substrate because it is multiply realizable. The term irreducible is IIT is used instead to indicate that a substrate's cause–effect power is not the same as that of its parts.

[xxxiv] Of course, there would be some difference in the substrate, namely a breakdown of causal interactions in the main complex owing to neuronal bistability (Tononi et al., 2024).

[xxxv] The duration of the macro state may however differ for different macro units belonging to the same complex.

[xxxvi] The multiple realizability claim has been qualified more and more stringently. Even so, a mental state is still identified with its causal role—what it does—rather than with a cause-effect structure—what it is.

[xxxvii] An additional issue is that computations and functions that we attempt to project into the brain's mechanisms to understand what the brain might be doing may only partially fit its evolutionary and developmental history.

[xxxviii] This account also suggests why other circuits within our brain, such as subcortical pathways mediating similar spatial functions, like foveation and tracking, may do so without contributing to experience: because these circuits are organized as connected maps (lacking lateral connections), rather than as connected grids (M. Grasso et al., 2021).

[xxxix] Berkeley's notion of 'esse est percipi' (to be is to be perceived) comes to mind here. However, while IIT recognizes that only the apple in the mind has intrinsic existence as a content of experience, so it genuinely exists, it does not deny the extrinsic existence of the apple on the table as a tight aggregate of ontological dust.

[xl] Note that, in IIT's ontology, there is no room for a 'Platonic' realm of ideal entities or concepts. Everything that genuinely exists, exists in the mind of a conscious subject, here and now. 'Platonic' ideas would be best characterized as concepts that are highly intersubjective and highly persistent.

[xli] A causal process can also be interpreted as a realization of the causal potential of an entity in a state over many transitions, given some background conditions.

[xlii] Physics often adopts the complementary approach of holistic prediction: if we have the equation predicting the system as a whole, given some initial conditions, we can predict each individual unit, or subset of units.

[xliii] Science trumps 'common sense' when it comes to extrinsic existence. But when it comes to intrinsic existence, common sense trumps mainstream science (to be precise, scientific approaches that deny or ignore the intrinsic perspective, unlike IIT's intrinsic ontology). Common sense says experience is fundamental, mainstream science says no (or it is a misnomer, an illusion, something that cannot be found objectively). Common sense says we have free will (it is I who just decided to raise my hand), mainstream science says no (or it is an illusion). As Dr. Johnson put it: "All theory is against the freedom of the will; all experience is for it."

[xliv] In this sense, the right question to ask about free will is not 'could I have decided otherwise?' but rather, 'was there an alternative in my mind?' If there was, and I decided based on values, purposes, and reasons, then I had free will.

[xlv] The causal account on the effect side of my decision would behave similarly, with the difference that, while my decision is a content within an intrinsic entity and thereby the actual cause of downstream occurrences (say, my motoneurons firing and my muscles contracting to press SEND), the action I perform is not: it only exists in my mind when I think of it as an 'action.'

[xlvi] Note that IIT's causal process analysis does not rule out 'unconscious' causation, say when, having decided to express my thinking, the words flow effortlessly, in the proper order and with the proper articulation, without my conscious intervention. In such cases, the causal account of each step in this process is likely to overlap with many small complexes, rather than with my main complex. The overall causal process can still be characterized extrinsically as a relative maximum and be further subdivided into many small causal streams ('rivulets'), each of them supported by entities that exist intrinsically, but minimally so.

[xlvii] If one could directly observe with a '$\Phi$-scope' or 'qualiascope' (Tononi, 2012) what genuinely exists when I decide of my own free will, one would not see any neurons (or atoms) at all, but only an immense cause-effect structure, renewing itself at every moment, which causes what happens next (say, my action). This visualization should help realizing that the neurons cannot cause anything because they do not even exist (intrinsically). Only my decision exists (and before that alternatives, values, goals, reasons, and intentions in my mind), therefore only my decision can be the cause of my action.

[xlviii] This conclusion does not mean that artificial consciousness is altogether impossible. A machine built of non-organic ingredients could support consciousness, much like certain parts of our brain can, as long as its substrate is maximally irreducible and its units are maximally irreducible within.

[xlix] By the same token, our brain can 'see' functionally without any phenomenal accompaniment. For example, the brain plans and executes saccades to functionally relevant features of a visual scene without our awareness multiple times a second (Pelisson et al., 1986) (see also (M. Grasso et al., 2021)).

[l] In our case, the end result seems to be that we can only be intelligent when we are conscious—when we are unconscious, we can hardly do anything at all. This suggests that the substrate of consciousness must be in working order to 'ping' cognitive and executive processes that mediate intelligent behaviors.

[li] In a way, such misreadings are to be expected. After all, according to IIT, all meaning is intrinsic, and perception is always an interpretation.

[lii] Aaronson proves his point in a mathematical context—that of expander graphs—but he correctly recognizes that IIT would require a physical implementation in grids with dense connectivity (fan-in and fan-out).